\documentclass[fleqn,usenatbib]{mnras}

\usepackage{newtxtext,newtxmath}

\usepackage[T1]{fontenc}

\DeclareRobustCommand{\VAN}[3]{#2}
\let\VANthebibliography\thebibliography
\def\thebibliography{\DeclareRobustCommand{\VAN}[3]{##3}\VANthebibliography}

\usepackage{graphicx}	
\usepackage{amsmath}	
\usepackage{multirow}

\RequirePackage{color}\definecolor{GREEN}{rgb}{0,0.45,0} 

\title[The JWST weather report from WISE 1049AB II]{The JWST weather report from the nearest brown dwarfs II: Consistent variability mechanisms over 7 months revealed by 1–14 $\mu$m NIRSpec + MIRI monitoring of WISE 1049AB}

\author[Xueqing Chen et al.]{Xueqing Chen,$^{1}$\thanks{E-mail: xueqing.chen@ed.ac.uk}
Beth A. Biller$^{1,2}$,
Xianyu Tan$^{3}$, 
Johanna M. Vos$^{4}$,
Yifan Zhou$^{5}$,  
Genaro Su\'arez$^{6}$, 
\newauthor  
Allison M. McCarthy$^{7}$,
Caroline V. Morley$^{8}$,
Niall Whiteford$^{6}$, 
Trent J. Dupuy$^{1,2}$, 
Jacqueline Faherty$^{6}$, 
\newauthor 
Ben J. Sutlieff$^{1,2}$, 
Natalia Oliveros-Gomez$^{9}$,
Elena Manjavacas$^{9, 10}$, 
Mary Anne Limbach$^{11}$,
Elspeth K. H. Lee$^{12}$
\newauthor
Theodora Karalidi$^{13}$,
Ian J.M. Crossfield$^{14}$, 
Pengyu Liu$^{1,2,15}$, 
Paul Molliere$^{16}$,
Philip S. Muirhead$^{7}$, 
\newauthor
Thomas Henning$^{16}$, 
Gregory Mace$^{8}$, 
Nicolas Crouzet$^{15}$, 
Tiffany Kataria$^{17}$
\\
$^{1}$ Institute for Astronomy, University of Edinburgh, Royal Observatory, Edinburgh EH9 3HJ,UK \\
$^{2}$ Centre for Exoplanet Science, University of Edinburgh, Edinburgh, UK \\
$^{3}$ Tsung-Dao Lee Institute \& School of Physics and Astronomy, Shanghai Jiao Tong University, Shanghai 201210, People’s Republic of China\\
$^{4}$ School of Physics, Trinity College Dublin, The University of Dublin, Dublin 2, Ireland \\
$^{5}$ Department of Astronomy, University of Virginia, 530 McCormick Rd., Charlottesville, VA 22904, USA \\
$^{6}$ Department of Astrophysics, American Museum of Natural History, Central Park West at 79th Street, NY 10024, USA \\
$^{7}$ Department of Astronomy \& The Institute for Astrophysical Research, Boston University, 725 Commonwealth Ave., Boston, MA 02215, USA \\
$^{8}$ Department of Astronomy, The University of Texas at Austin, Austin, TX 78712, USA \\
$^{9}$ William H. Miller III Department of Physics and Astronomy, Johns Hopkins University, Baltimore, MD 21218, USA \\
$^{10}$ AURA for the European Space Agency (ESA), ESA Office, Space Telescope Science Institute, 3700 San Martin Drive, Baltimore, MD, 21218 USA \\
$^{11}$ Department of Astronomy, University of Michigan, Ann Arbor, MI 48109, USA \\
$^{12}$ Center for Space and Habitability, University of Bern, Bern, Switzerland\\
$^{13}$ Department of Physics, University of Central Florida, 4111 Libra Dr., Orlando, FL 32816 \\
$^{14}$ Department of Physics and Astronomy, University of Kansas, Lawrence, KS, USA \\
$^{15}$ Leiden Observatory, Leiden University, PO Box 9513, 2300 RA Leiden, The Netherlands \\
$^{16}$ Max-Planck-Institut f\"ur Astronomie, K\"onigstuhl 17, D-69117 Heidelberg Germany \\
$^{17}$ Jet Propulsion Laboratory, California Institute of Technology, Pasadena, CA 91109 USA \\
}

\date{Accepted XXX. Received YYY; in original form ZZZ}

\pubyear{\the\year{}}

\begin{document}
\label{firstpage}
\pagerange{\pageref{firstpage}--\pageref{lastpage}}
\maketitle

\begin{abstract}
We present a new epoch of {\sl JWST} spectroscopic variability monitoring of the benchmark binary brown dwarf WISE 1049AB, the closest, brightest brown dwarfs known. Our 8-hour MIRI low resolution spectroscopy (LRS) and 7-hour NIRSpec prism observations extended variability measurements for any brown dwarfs beyond 11~$\mu$m for the first time, reaching up to 14~$\mu$m. Combined with the previous epoch in 2023, they set the longest {\sl JWST} weather monitoring baseline to date.
We found that both WISE 1049AB show wavelength-dependent light curve behaviours. Using a robust k-means clustering algorithm, we identified several clusters of variability behaviours associated with three distinct pressure levels. By comparing to a general circulation model (GCM), we identified the possible mechanisms that drive the variability at these pressure levels:
Patchy clouds rotating in and out of view likely shaped the dramatic light curves in the deepest layers between 1--2.5~$\mu$m, whereas hot spots arising from temperature / chemical variations of molecular species likely dominate the high-altitude levels between 2.5--3.6~$\mu$m and 4.3--8.5~$\mu$m. Small-grain silicates potentially contributed to the variability of WISE 1049A at 8.5-11~$\mu$m. 
While distinct atmospheric layers are governed by different mechanisms, we confirmed for the first time that each variability mechanism remains consistent within its layer over the long term.
Future multi-period observations will further test the stability of variability mechanisms on this binary, and expanded {\sl JWST} variability surveys across the L-T-Y sequence will allow us to trace and understand variability mechanisms across a wider population of brown dwarfs and planetary-mass objects.

\end{abstract}




\section{Introduction}
The James Webb Space Telescope ({\sl JWST}) is transforming the field of time-resolved observations of brown dwarfs and planetary-mass objects. Occupying the mass range between the heaviest gas giant planets and the lightest stars, brown dwarfs offer a unique window to study atmospheric processes in ultracool environments. As fast rotators with rotational timescales of hours, they enable studies using time-resolved observations to capture the inhomogeneity of their atmospheres and reveal the underlying mechanisms driving their weather.

Previous photometric and narrow-band spectroscopic monitoring has shown that variability is common in brown dwarfs and planetary-mass objects, in the optical ($<$1~$\mu$m; \citealt{Heinze2015, Apai2021}), near-IR (1--$\sim$2.5 $\mu$m; \citealt{Radigan2012, Radigan2014, Vos2019, Lew2020, Liu2023}) and mid-IR ($>$3$\mu$m; \citealt{Metchev2015, Vos2020, Vos2022}).
Proposed mechanisms for this observed variability include patchy silicate clouds \citep{Marley2010, Apai2013, Tan2021, Vos2023}, hot spots \citep{Robinson2014}, non-equilibrium chemistry \citep{Tremblin2016, Tremblin2020, Lee2024}, and aurorae \citep{Hallinan2015, Kao2016, Faherty2024}.

Past variability studies have left a few unsolved puzzles: 
Phase shifts in the peaks and troughs of the light curves or even completely different light curve shapes has been found in different narrow bands, e.g. in the near vs. mid IR, pointing to different atmospheric structures and mechanisms in distinct vertical layers of the atmosphere \citep{Buenzli2012, Yang2016, Karalidi2016, Biller2018, Biller2024}. 
Moreover, variability can evolve vigorously over years-long timescales \citep{Apai2021, Zhou2022, Fuda2024}, and in some cases it can be irregular and even change dramatically on period-to-period timescales \citep{Biller2024}. It remains unclear whether these changes were a result of different variability mechanisms caught at different epochs, or just chaotic appearances of the same mechanisms. 

Because of the highly complex and dynamic nature of these atmospheres, time-resolved spectroscopic observations with a wide simultaneous wavelength coverage are required to capture various effects at once: a range of molecular absorption features that are sensitive to different pressures, as well as the continuum which is sensitive to clouds. This approach allows us to obtain a comprehensive 3-D view of the atmosphere and disentangle the variability-driving mechanisms at all atmospheric depths \citep{Morley2014, Biller2024, McCarthy2024}.
Moreover, repeated and multi-epoch observations of such are essential to address the rapid evolution of their atmospheric behaviours.
{\sl JWST} serves as a unique tool for this task with its unprecedented sensitivity, spectral coverage, and resolution, with the NIRSpec instrument covering 0.6–5 $\mu$m \citep{Jakobsen2022}, while MIRI extending this coverage to 28 $\mu$m.


Recently, {\sl JWST} has enabled multi-period monitoring of the benchmark binary brown dwarf system WISE~J104915.57–531906.1AB (also known as Luhman~16AB, \cite{Luhman2013}; hereafter WISE 1049AB) \citep{Biller2024}. 
With a distance of 1.9960$\pm$0.0002 pc, they are the closest and brightest brown dwarfs to Earth \citep{Bedin2024}.
Both have dynamical mass measurements (34.2$\pm$1.2 M$_{\rm{Jup}}$ for A, 27.9$\pm$1.0 M$_{\rm{Jup}}$ for B, \citealt{Bedin2017, Garcia2017, Lazorenko2018}), with a mass ratio of $\sim$0.82. The system has a well-measured age of 510$\pm$95 Myr via membership in the newly discovered Oceanus moving group \citep{Gagne2023}.
The A and B components span the L/T transition, with spectral types L7.5$\pm$1 and T0.5$\pm$1 respectively \citep{Burgasser2013}. Since both A and B appear to be typical L/T transition objects given their spectra, they offer a unique case for studying heterogeneous atmospheric structures and comparing different phases in the L/T transition, which is currently poorly understood.

\begin{figure*}
\includegraphics[width=\textwidth]{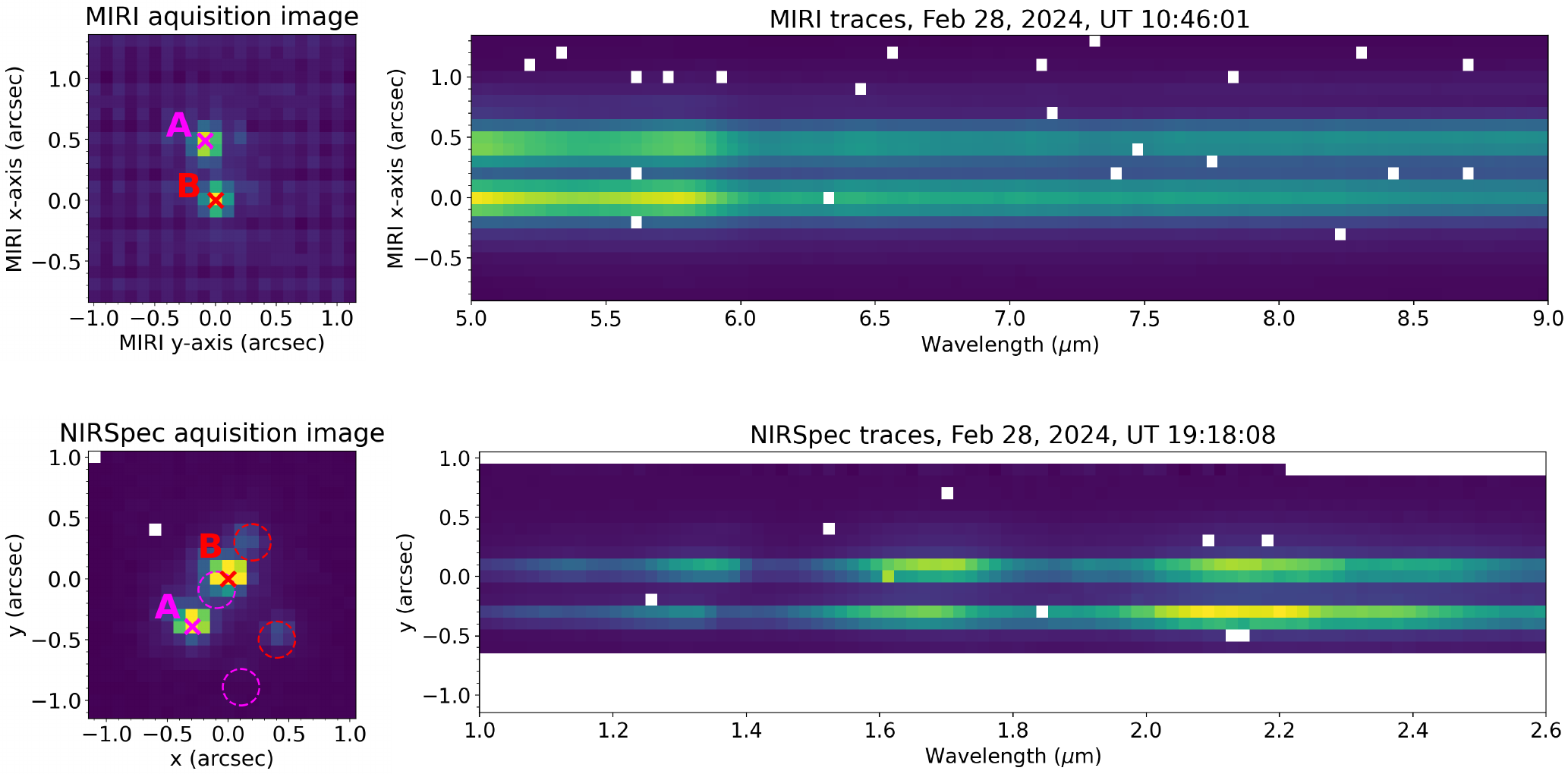}
\caption{On the left: MIRI and NIRSpec target acquisition images showing the resolved point sources for WISE 1049 A (magenta) and B (red). The NIRSpec acquisition image also shows the positions of the additional point sources (dotted circles) caused by the mirror tilt event predicted by simulation using \texttt{STPSF} (formerly \texttt{WebbPSF}).
On the right: {\sl JWST} pipeline stage 2 calint images showing cropped regions of the resolved traces of WISE 1049AB from MIRI and NIRSpec. The detector pixels and hence the position of the two components are horizontally aligned with the acquisition images on the left. The white pixels indicate bad pixels on the MIRI/NIRSpec detector.}
\label{fig:obs}
\end{figure*}

Given their proximity and brightness, extensive studies have sought to characterize their variability at optical, near-IR and mid-IR wavelengths 
(e.g., \citealt{Gillon2013, Biller2013, Burgasser2014, Mancini2015, Street2015, Buenzli2015a, Buenzli2015, Kellogg2017, Bedin2017, Apai2021, Heinze2021, Fuda2024}). Both components are found to be variable, with WISE 1049B being the main source of variability of the binary in optical to near-IR, with amplitudes of 5--11\% \citep{Biller2013, Gillon2013, Buenzli2015}, whereas WISE 1049A has a maximum measured variability of only $\sim$4\% at 0.8--1.15 $\mu$m \citep{Buenzli2015}. Variability amplitudes for both components can vary notably from one rotational period to another \citep[e.g.][]{Apai2021, Fuda2024, Biller2024}, suggesting dramatic atmospheric evolution even on rotational timescales. 
A few studies have shown evidence for patchy clouds \citep{Crossfield2014, Buenzli2015a, Karalidi2016, Chen2024} or planetary-scale storms/waves \citep{Apai2021, Fuda2024} in shaping the variability on WISE 1049B. 
Before {\sl JWST}, no study could directly test the variability caused by silicate clouds at their $\sim$9 $\mu$m absorption feature \citep{Luna2021}. \cite{Biller2024} reported the first {\sl JWST} NIRSpec + MIRI monitoring for WISE 1049AB from 1--11 $\mu$m, and no resolved monitoring has been reported for these two objects beyond 11 $\mu$m. \cite{Biller2024} identified three distinctive groups of light curve behaviours, with the transitions between them aligning with molecular absorption bands, particularly the CO band at $\sim$4.2~$\mu$m and the silicates band at $\sim$8.5~$\mu$m. However, the MIRI traces in their data are separated by only $\sim$2 pixels on the detector and must be disentangled via point spread function (PSF) fitting, resulting in low signal-to-noise ratio (SNR) beyond 10~$\mu$m and preventing reliable studies of the variability at silicate features around 8.5--11 $\mu$m. 

In this second paper of the series, we present a new epoch of spectroscopic variability monitoring of WISE 1049B which is spatially resolved from 1--14 $\mu$m. In combination with the previous epoch of observation in paper I \citep{Biller2024}, this provides a robust test of the long-term variability behaviours of this benchmark binary.
We describe our observation and data reduction methods in Sections \ref{sec:obs} and \ref{sec:reduction}. The spectra and light curves are presented in Section \ref{sec:results}, followed by analysis and discussion in Section \ref{sec:analysis}, including light curve clustering, interpretation with atmospheric models, and comparison with a general circulation model (GCM). Finally, we summarize our findings in Section \ref{sec:conclusion}.


\section{Observations}
\label{sec:obs}

As part of {\sl JWST} GO 2965 (PI: Biller), we observed at least one full period of both the WISE 1049A and B components with MIRI Low Resolution Spectroscopy (LRS) and NIRSpec PRISM respectively in a consecutive sequence in order to capture similar variability properties with both instruments. 
MIRI LRS time-series observations (TSOs) were acquired from UT 10:46:01 to UT 19:18:04 on Feb 28, 2024.
These observations were directly followed by NIRSpec Bright Object Time Series (BOTS) observations from UT 19:18:08 on 2024 Feb 28 to UT 03:06:23 on 2024 Feb 29. The observing strategy was the same as the first epoch taken in July 2023 \citep{Biller2024}, except that the MIRI background observation was conducted before the MIRI LRS time-series instead of after it.

We obtained MIRI TSOs of WISE 1049AB with the LRS slitless mode with the P750L disperser to prevent slit losses, disabling dithers to ensure photometric stability, and using the FAST readout mode to provide many samples up the ramp. This yields spectral resolutions of R = 40–160 from 5 to 14 $\mu$m.
We performed target acquisition using the LRS SLITLESSPRISM subarray with the FAST readout pattern with four groups, followed by 2193 integrations ($\sim$8h) using the LRS SLITLESSPRISM subarray with the FASTR1 readout pattern with 80 groups per integration, providing a base exposure time of 12.72 s during the TSOs. 
The MIRI TSOs were preceded by a brief background observation from Feb 28, 2024 UT 09:22:09 to Feb 28, 2024 UT 10:45:57 of an empty field offset from the target by -10 arcsec, respectively, in right ascension and declination. We obtained 10 integrations with the same FASTR1 readout pattern and 80 groups per integration at the background position.

Then, we observed WISE 1049AB using the NIRSpec BOTS mode in the low-resolution PRISM/CLEAR mode, providing simultaneous wavelength coverage from 0.6 to 5.3 $\mu$m at resolutions from R = 30 to 300.
Adopting the SUB512S subarray, NRSRAPID readout, and 2 groups per integration, yielded a cadence of 0.45 s and avoided saturation of either component.

WISE 1049AB was observed at a V3 PA (position angle of the observatory V3 reference axis) of 356$^\circ$, corresponding to a MIRI aperture position angle (APA) of 0.8$^\circ$ and a NIRSpec APA of 137.2$^\circ$. During the epoch of the {\sl JWST} observation, WISE 1049B was at a separation of $\sim$0.5 arcsec and a position angle of 101$^\circ$ from WISE 1049A, as measured from the NIRSpec acquisition and MIRI target verification images. As a result, the NIRSpec and MIRI traces are both well resolved. The NIRSpec traces are separated by $\sim$ 4 pixels, and the MIRI traces are separated by $\sim$ 4.5 pixels, as shown in the right-hand panels of Fig. \ref{fig:obs}.

\subsection{Effect of the mirror tilt event}

Our data was affected by the largest mirror “tilt
event” since the commissioning of {\sl JWST} shortly before the observations. Tilt events are occasional abrupt shifts in the position of one or more mirror segments thought to arise from structural micro-dynamics within the telescope \citep{Rigby2023}. The wavefront stability is measured with NIRCam approximately every two days \citep{Acton2012, McElwain2023}. On UT 2024 February 27, NIRCam recorded a total wavefront error (WFE) of 527 nm RMS, significantly exceeding the baseline value of $<$70 nm RMS. This measurement represents the highest WFE observed since the commissioning of {\sl JWST}.

Our observations coincided with the mirror tilt event, which notably affected the data, particularly the NIRSpec observations. In the target acquisition image, the tilt event introduced additional copies of the target PSFs.
To identify the origin and brightness of artificial point sources in the NIRSpec acquisition image introduced by the tilt event, we simulated the degraded NIRSpec PSF using Optical Path Difference (OPD) files during the tilt event using \texttt{STPSF} (formerly \texttt{WebbPSF}).
Fig. \ref{fig:obs} shows the predicted positions of extra point sources on the NIRSpec target acquisition image by the simulation. 
We identified that the tilt event has caused each of WISE 1049A and B to have two extra images on the detector at 3\% of the original PSF flux level, one to the upper right of the original PSF and one to the lower right. The lower right copy of the B component falls almost directly to the right of the A component (less than 1 pixel in the $y$-direction). Because NIRSpec disperses along the $x$-direction, this means that the trace of WISE 1049A is contaminated by this extra copy of WISE 1049B at $\sim$3\% level. The upper right copy of the A component falls within 2 pixels of the original position of B. This means that the trace of WISE 1049B is also contaminated by the extra copy of A, although less severely. Because the traces of A and B components start at different $x$ positions on the detector, this means that the blended A and B traces also have different wavelength scales. In effect, this adds a slightly shifted version of B's spectra to A with a flux ratio of 0.03 and vice versa. This source of contamination should be kept in mind while interpreting the NIRSpec spectra and light curves in the later results section.

In comparison, the MIRI PSF was also degraded, but there was no blending issue. The simulated degraded MIRI PSF shows wider FWHM, but there are no extra copies of the original PSF created. Given that the MIRI traces were well separated in this epoch, we believe that the effect of the mirror tilt event on the MIRI observations was not significant.

\section{Data reduction and spectral extraction}
\label{sec:reduction}

\begin{figure*}
\centering
\begin{minipage}{0.49\textwidth}
  \centering
  \includegraphics[width=1\linewidth]{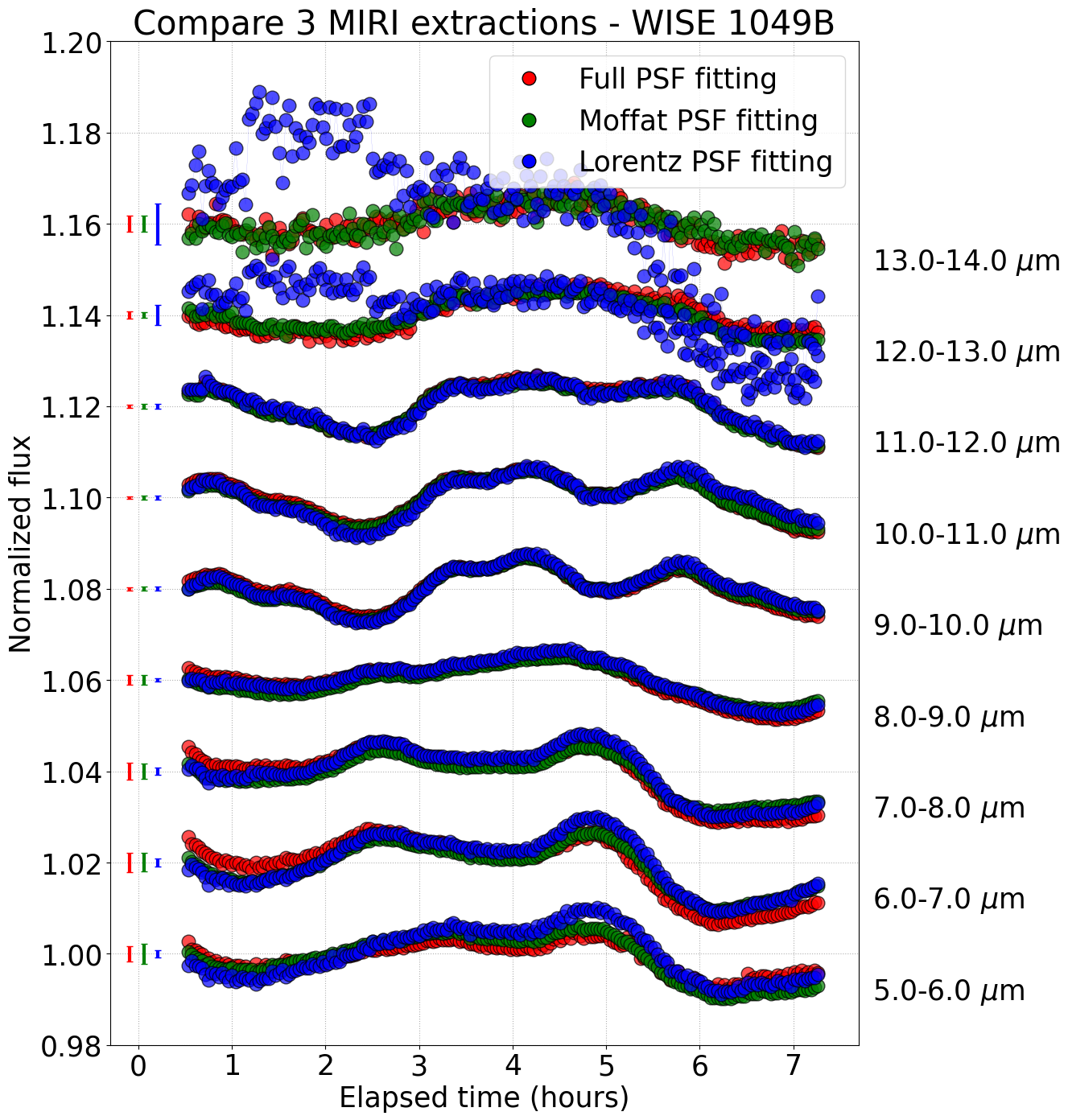}
\end{minipage}
\begin{minipage}{0.49\textwidth}
  \centering
  \includegraphics[width=1\linewidth]{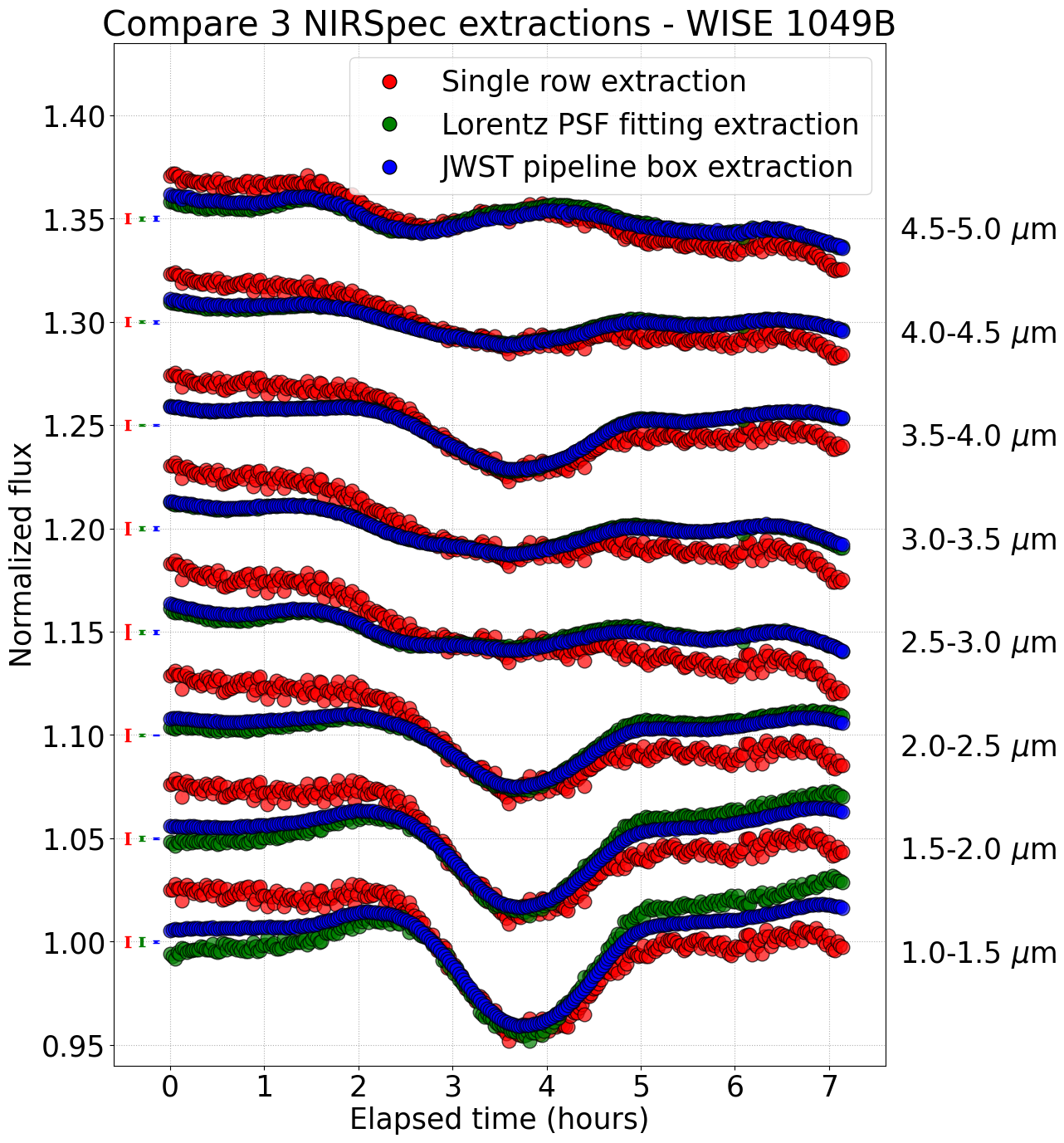}
\end{minipage}
\caption{Sample WISE 1049B light curves from the different spectral extraction methods. Left: The 3 different MIRI extractions -- Full commissioning PSF fitting (red), Moffat profile PSF fitting (green), and Lorentz profile PSF fitting (blue). Right: The 3 different NIRSpec extractions -- Single row extractions where we took row 3 in the {\sl JWST} pipeline stage 2 calint images as the spectra for WISE 1049A and row 7 for WISE 1049B (red), Lorentz profile PSF fitting (green), and standard {\sl JWST} pipeline extraction (blue). The error bars for the light curves are estimated by $\sigma_\mathrm{pt}$ \citep{Biller2024} and are shown on the left of each light curve.}
\label{fig:reduction_comparison}
\end{figure*}

\subsection{MIRI reduction and spectral extraction}

We performed basic data reduction steps on the LRS time series using the STScI {\sl JWST} pipeline (version 1.13.4, CRDS version 11.17.14, and CRDS context file 'jwst\_1210.pmap'). The uncalibrated files for the LRS TSO were divided into 11 segments due to the length of the observation. We ran all segments of data through pipeline stages 1 and 2 (\texttt{calwebb detector1} and \texttt{calwebb spec2}) with default settings. From the resulting calibrated 2D spectral image (calints) products, we then used custom scripts to extract the spectra of both components of the binary using PSF fitting techniques. 

In the LRS slitless spectroscopy mode, the spectral traces for the A and B components of the binary are dispersed in the $y$-direction starting from different $y$-positions due to the offset in $y$ between the two components. Thus, we need to correct the wavelength solution for each component to account for their offsets from the nominal pointing position. 
In the MIRI acquisition images, we found that the B component was placed at the nominal pointing position. Using \texttt{astropy.photutils.DaoStarfinder}, we measured A being offset from B by 4.46 pixels (0.49 arcsec) in $x$ and by 0.78 pixels (0.08 arcsec) in $y$.
To correct for this offset, we interpolated the wavelength solution from 1D extraction from the {\sl JWST} pipeline for A. Because of this offset, the first wavelength bin in the MIRI spectrum for WISE 1049B is at 4.17 $\mu$m, while the first wavelength bin for WISE 1049A is at 4.34 $\mu$m. 

To robustly extract the 1D spectra from the calints images, we tried three different PSF-fitting extraction methods, following the details in \cite{Biller2024}:
1) \textit{Full commissioning PSF fitting} -- to fit the MIRI PSF model provided by STScl (S. Kendrew and G. Sloan, private communication, STScI) at each wavelength slice with the positions and amplitudes for A and B components as free parameters; 
2) \textit{Lorentzian profile PSF fitting} -- to fit two Lorentzian profiles at each wavelength slice with the positions, FWHMs, and amplitudes of both components as free parameters; and
3) \textit{Moffat profile PSF fitting} -- Similar to Lorentzian fitting but using a Moffat profile, as defined in \cite{Biller2024}. 
In all of the above methods, a slice of sky region was used to estimate the background level, which is subtracted from the traces before fitting a PSF model. 

Because the MIRI traces are well separated in this epoch, we were able to achieve reliable extraction up to 14 $\mu$m, making this the longest wavelength variability monitoring of any brown dwarf or planetary-mass object to date.
The left-hand panel of Fig. \ref{fig:reduction_comparison} compares the sample light curves in 1 $\mu$m wavelength bins produced by all three extraction methods for WISE 1049B. The three extractions agree with each other except for the Lorentz PSF-fitting in the longest wavelength bin, where the fitting yields low SNR $>$12 $\mu$m.
Considering the extraction quality at all wavelengths, we adopt the full PSF fitting extraction for the rest of the MIRI analysis. The error bars of the extraction are calculated from the covariance matrices of the fitted amplitude parameter for WISE 1049 A and B.

\subsection{NIRSpec reduction and spectral extraction}

For the NIRSpec observations, we first performed basic data reduction using the {\sl JWST} pipeline (version 1.18.0, CRDS version 12.1.4, and CRDS context file 'jwst\_1364.pmap'). The NIRSpec BOTS uncalibrated files were divided into 5 segments. We process all segments through pipeline stages 1 and 2 (\texttt{calwebb detector1} and \texttt{calwebb spec2}) with default settings except for the corrections detailed below. 

In the NIRSpec target acquisition image, the B component of the binary was acquired at the nominal pointing position, and the A component is on the left to it on the detector.
Using \texttt{astropy.photutils.DaoStarfinder} we measured the A component to be offset from B by -2.94 pixels (-0.29 arcsec) in the $x$-direction and by -3.91 pixels (-0.39 arcsec) in the $y$-direction. Because NIRSpec disperses along the $x$-axis, this means the wavelength solution for the A component will start with an offset from B.
To make the pipeline automatically correct for this offset at the wavelength correction (\texttt{wavecorr}) step in stage 2, we set \texttt{XOFFSET} = -0.29 and \texttt{YOFFSET} = -0.39 in the fits file headers for the A component before running pipeline stage 2. We also set \texttt{TSOVISIT = False} in the file headers as the TSO version of the \texttt{spec2} pipeline skips the \texttt{wavecorr} step. The pipeline then treats the offset as an additional dither position and adjusts the wavelengths for the A component.

In the NIRSpec observation, the two components of the binary are generally well resolved, but each component is contaminated by the other component due to the extra PSFs resulting from the tilt event. 
Therefore, in addition to the standard spectral extraction from the \texttt{extract\_1d} step in stage 2, we tested two other approaches to extract the spectra of both components of the binary from the resulting calibrated 2D spectral image (calints), aiming to mitigate the effects of the tilt event. We describe all three extraction approaches below:

1) \textit{JWST pipeline extraction}. The \texttt{extract\_1d} step in the STScI {\sl JWST} pipeline essentially sums up the flux in a user-selected box region with background subtraction.
We override the default extraction region using an updated JSON file to use a rectangular extraction region centered on each component (rows 0--4 for A and row 5--9 for B), with a background subtraction region placed in a source-free region at the bottom of the detector (rows 11--14). 

2) \textit{Lorentzian PSF fitting extraction}.
We tested with PSF-fitting extractions to better account for the potentially degraded PSF widths from the tilt event. Using the same double Lorentzian model as in the MIRI data extraction, we fit each 2D spectral image in the NIRSpec data at each wavelength column. We allow the profile center (pixels in $y$-direction), profile amplitude, and profile FWHM (in pixels) for both components to be free parameters. To remove the small jitters in the fitted $y$-position, we performed an initial fit and then smoothed the $y$-positions using a 3rd-order polynomial interpolation. A re-fit was then performed with the $y$-positions fixed to the polynomial.

3) \textit{Single-row extraction}. 
Finally, we performed a row-by-row analysis on the NIRSpec images to see which rows were less affected by the contamination from the other components. We took each row of the calints trace image as an individual spectrum and constructed light curves from the time-resolved spectral arrays for each row. The row-by-row light curves can be found in Appendix \ref{fig:rowbyrowlc}. 
Because the amount of contamination from the other component is different in each row and the contamination is wavelength-dependent, the resulting light curves from different rows showed different shapes. When matching the rows of the NIRSpec acquisition image in Fig. \ref{fig:obs} to the trace image, we found that row 3 and row 7 represent the least contaminated spectra for WISE 1049A and B respectively. We therefore take the single-row spectra from row 3 and row 7 to represent the spectra of WISE 1049A and B.

The right-hand panel of Fig. \ref{fig:reduction_comparison} compares the sample NIRSpec light curves in 0.5 $\mu$m wavelength bins constructed from the spectra from the three extraction methods described above for WISE 1049B. The three extractions agree with each other in general, with single-row extractions having an additional overall trend of decreasing brightness over time, and Lorentzian PSF-fitting yielding low SNR at the shortest wavelength bin. For the rest of the analysis, we will use extraction from the {\sl JWST} pipeline. The error bars of the final extraction are read from the x1dints files from the stage 2 pipeline.

\begin{figure*}
\includegraphics[width=\textwidth]{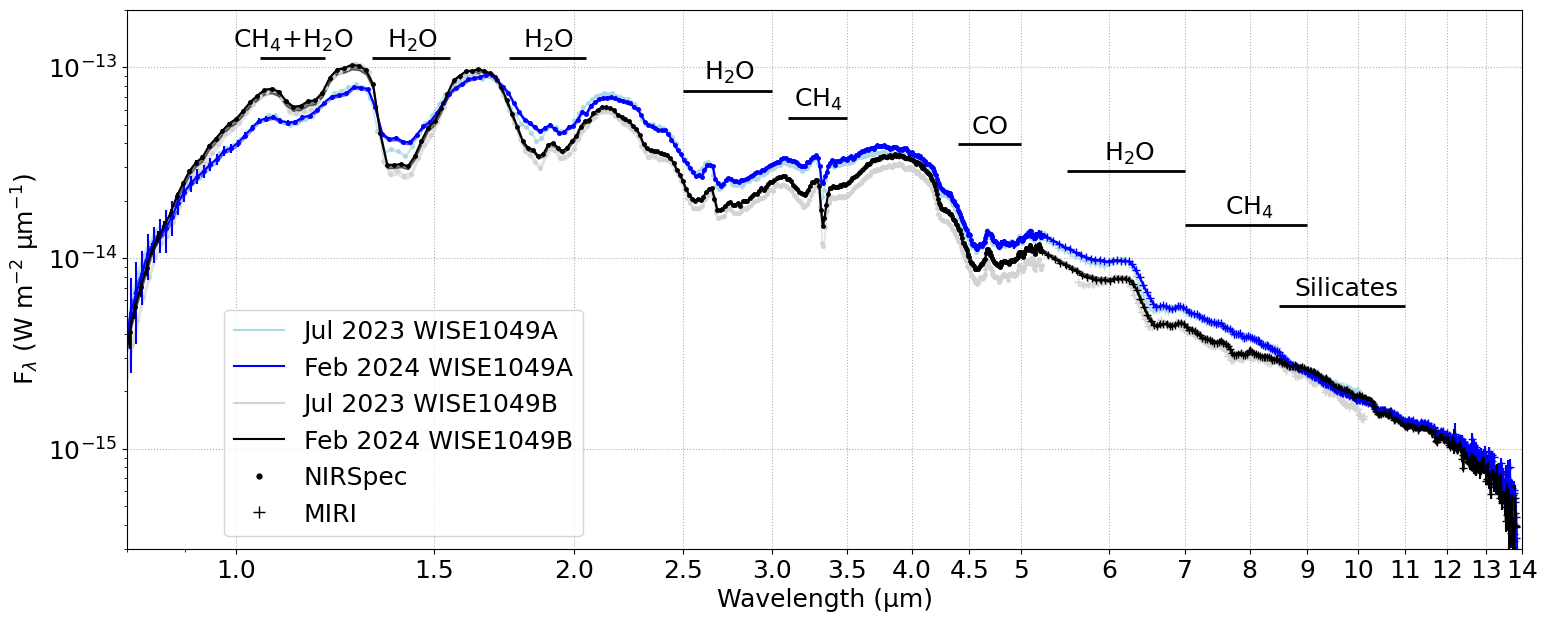}
\caption{The NIRSpec PRISM + MIRI LRS spectra for WISE 1049A (blue) and B (black). Dot and plus markers depict the mean flux across the whole observation period, while shaded areas show the spread of the spectra between the minimum and maximum spectrum. The spectra from the Feb 2024 epoch (presented in this work) are shown in dark blue and black, while the spectra from the July 2023 epoch \citep{Biller2024} are shown in light blue and grey in the background for comparison. Main spectral absorption features are labeled on top of the spectra. The errors are shown as vertical bars over each data point.
The error bars are smaller than the markers and therefore not visible, except for the regions $<$1$\mu$m for Feb 2024 WISE 1049A. This increase of error at the shortest wavelength for WISE 1049A could be caused by the contamination from the mirror tilt event.}
\label{fig:spectra}
\end{figure*}

\begin{figure*}
\includegraphics[width=0.85\textwidth]{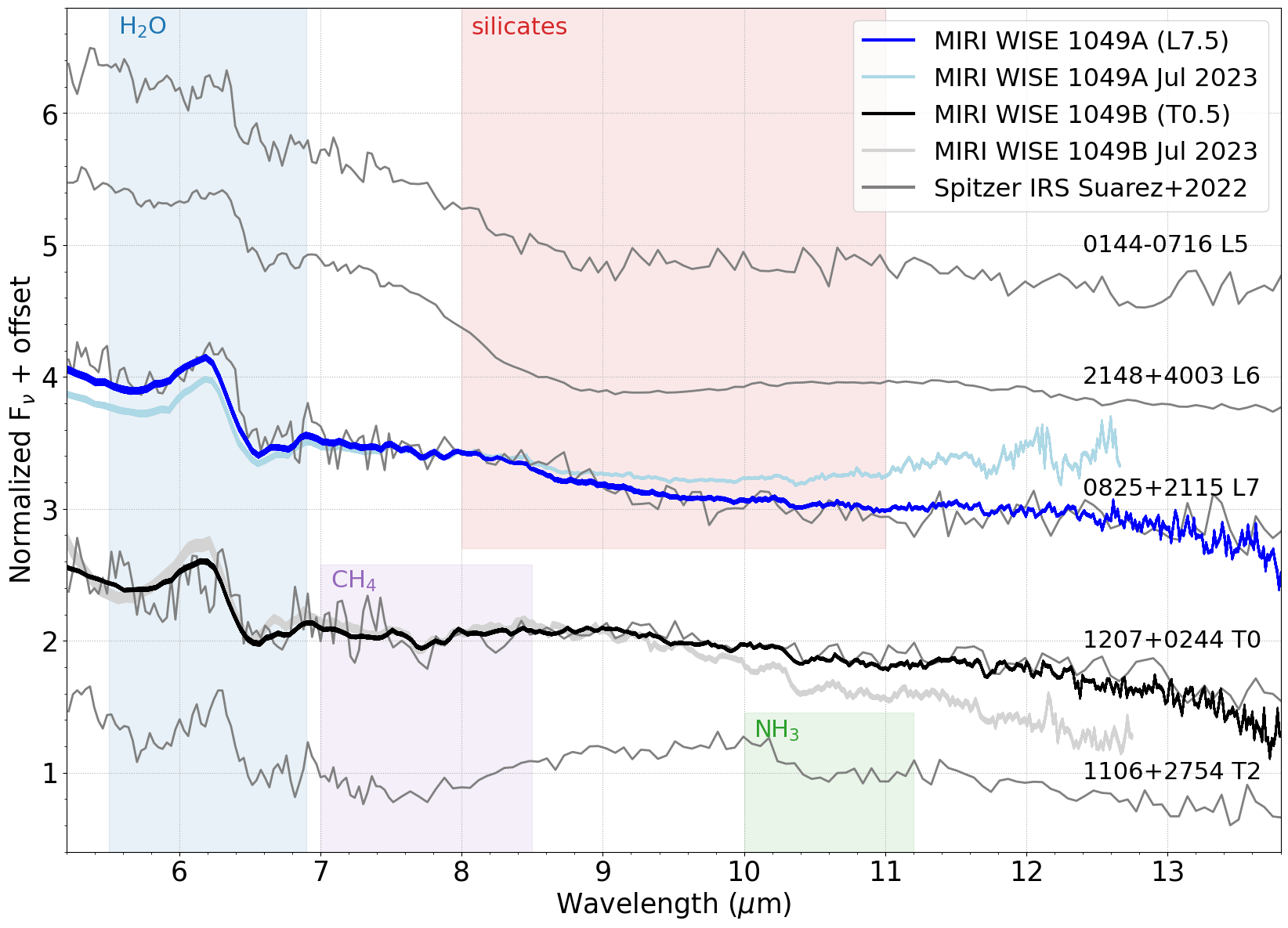}
\caption{Zoomed-in MIRI LRS spectra for WISE 1049A (blue) and B (black) in $F_{\nu}$ units, normalized with respect to their median value and over-plotted with Spitzer IRS spectra of five L5--T2 brown dwarfs (grey) taken from \citet{Suarez2022}. The spectra of WISE 1049 A and B are shifted vertically to best match the Spitzer spectra of a brown dwarf of same spectral type (L7 and T0 respectively) in the background for comparison. The dominant spectral bands for the relevant spectral types are highlighted with shaded areas.}
\label{fig:miri_spectra}
\end{figure*}

\section{Results}
\label{sec:results}

\subsection{Mean spectra}

In Fig. \ref{fig:spectra} we present the final flux-calibrated MIRI LRS and NIRSpec PRISM spectra for WISE 1049 A and B. We obtained 2193 spectra over the $\sim$8 h observation of MIRI, and 57000 spectra over the $\sim$7 h of NIRSpec. For both instruments, we plot the mean spectrum over these time points (dot markers for NIRSpec and plus markers for MIRI) and we show the spread of the spectra over time with a shaded area. We also show the spectra from the July 2023 epoch presented in \cite{Biller2024} for comparison. Both WISE 1049A and B spectra are generally consistent with the July 2023 epoch, with some minor differences observed -- mainly, the spectrum of the B component appears slightly brighter than in the July 2023 epoch at the long-wavelength edge of NIRSpec. We see absorption features from H$_2$O, CO, and CH$_4$. The CH$_4$ absorption feature around 3.35 $\mu$m is present in both the late-L type WISE 1049A and early-T type WISE 1049B. 

In Fig. \ref{fig:miri_spectra} we show a zoomed-in view of the MIRI spectra, in comparison with the Spitzer IRS spectra of five L5--T2 brown dwarfs from \cite{Suarez2022}. We display the MIRI spectra in $F_\nu$ units as this better displays the structure in the spectra at longer wavelengths.
Compared with the July 2023 epoch, we were able to extend the extraction to near 14 $\mu$m using PSF fitting due to better separation of the MIRI traces at this epoch. Nevertheless, beyond wavelengths of 12 $\mu$m, there is a noticeable drop in SNR which causes the increased spread of the spectral points. 

The Spitzer IRS spectra of the L5 and L6 brown dwarfs from \cite{Suarez2022} clearly show a dip between 8--11 $\mu$m at the small-grain ($<$1 $\mu$m) silicate absorption feature. For WISE 1049A, this silicate feature is weakly detected, consistent with its late-L spectral type. 
The spectra of WISE 1049B does not exhibit this feature, but instead shows a weak methane absorption at 7--9 $\mu$m. This is consistent with the sedimentation of silicate clouds below the photosphere and disappearance of the silicate absorption past L8 type, as well as the onset of methane absorption for T spectral types, as reported by \cite{Suarez2022}.  
Note that the flux increase beyond 9 $\mu$m in the July 2023 epoch (light blue curve in Fig. \ref{fig:miri_spectra}) is likely an artifact due to the two components being blended. As the two are much better separated in this epoch, the upward trend is no longer observed. Since this upward trend begins just inside the silicate absorption feature, it could introduce a bias when measuring the depth of this feature in the 2023 epoch.

\begin{figure*}
\centering
\includegraphics[width=1\linewidth]{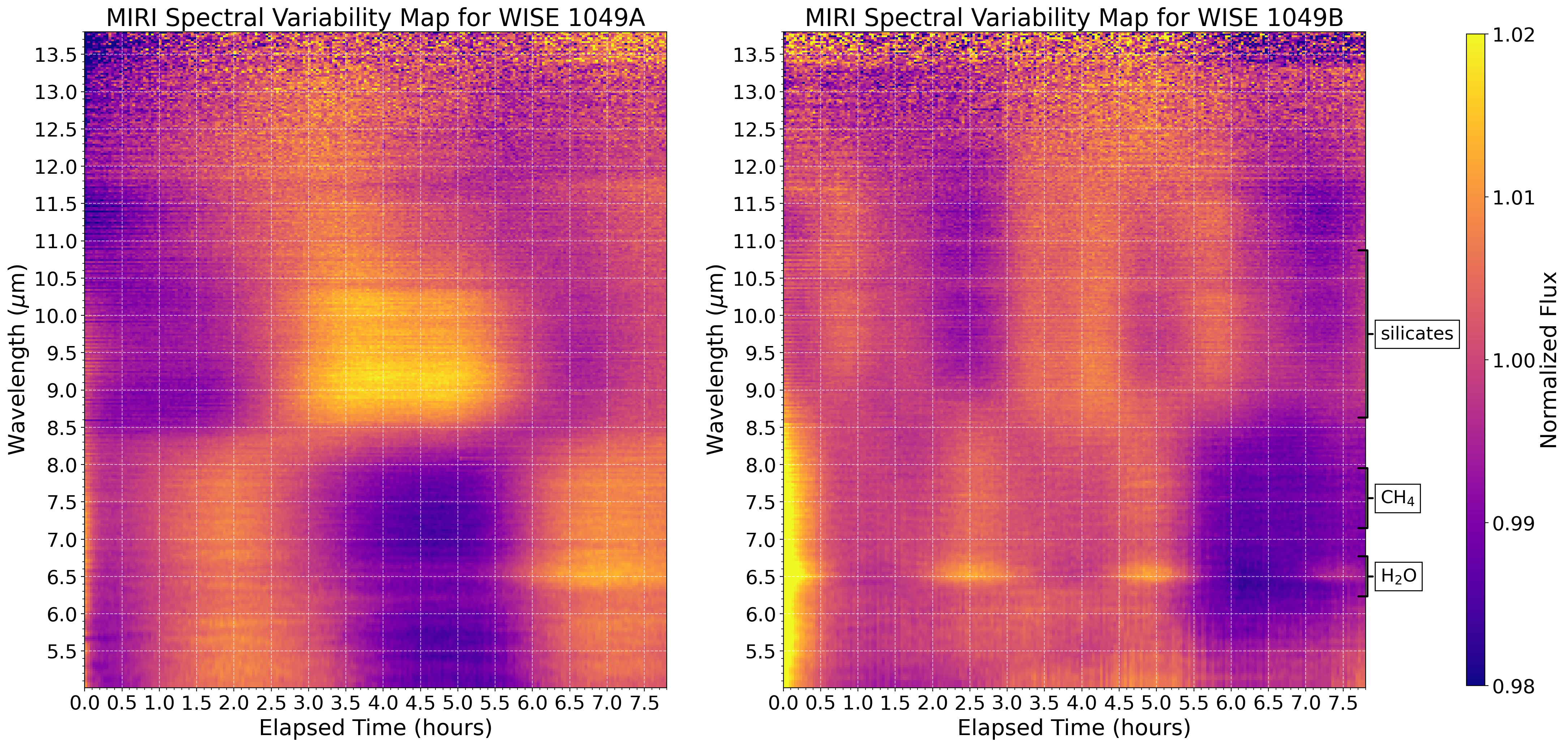}
\caption{Variability maps (spectroscopic light curves) of WISE 1049 A and B from MIRI, generated by producing median-normalized light curves for each wavelength bin in the MIRI spectra, then plotting as a function of time and wavelength. The colorbars range from fractional flux values of 0.98 to 1.02 to highlight wavelength-dependent changes in amplitude. The wavelengths of various molecular absorption bands are marked on the left of the maps. }
\label{fig:varmap_miri}
\end{figure*}

\begin{figure*}
\centering
\includegraphics[width=1\linewidth]{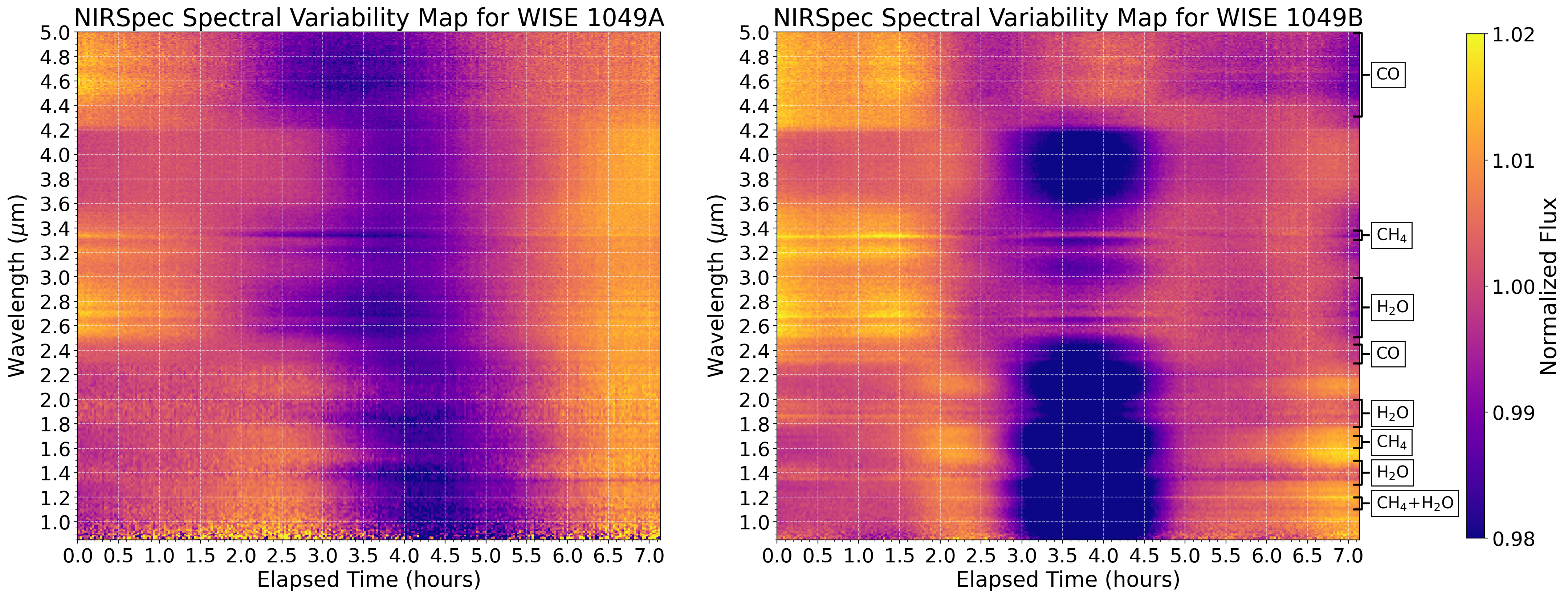}
\caption{Variability map (spectroscopic light curves) of WISE 1049 A and B from NIRSpec. Same as \ref{fig:varmap_miri} but from NIRSpec.}
\label{fig:varmap_nrs}
\end{figure*}

\begin{figure*}
\includegraphics[width=1\textwidth]{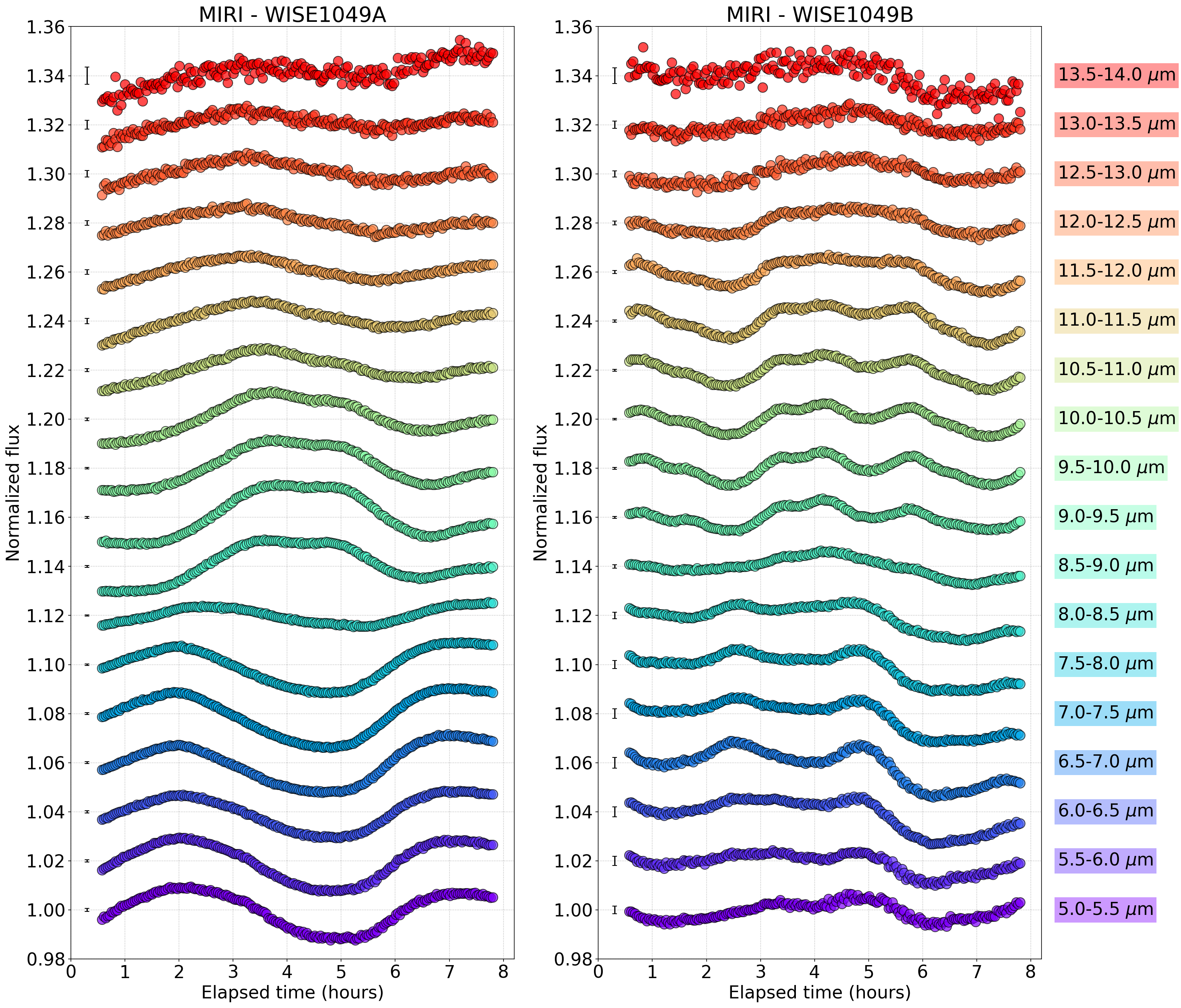}
\caption{MIRI light curves for WISE 1049A (left) and WISE 1049B (right), with 0.5 $\mu$m wavelength bins and a cadence of 129 s. Light curves have been normalized to their median values to highlight fractional variations. The light curves are offset with constants to visually separate them on the plot. The uncertainty for each light curve is shown as an error bar on the left side of the plot.}
\label{fig:lc_miri}
\end{figure*}

\begin{figure*}
\includegraphics[width=1\textwidth]{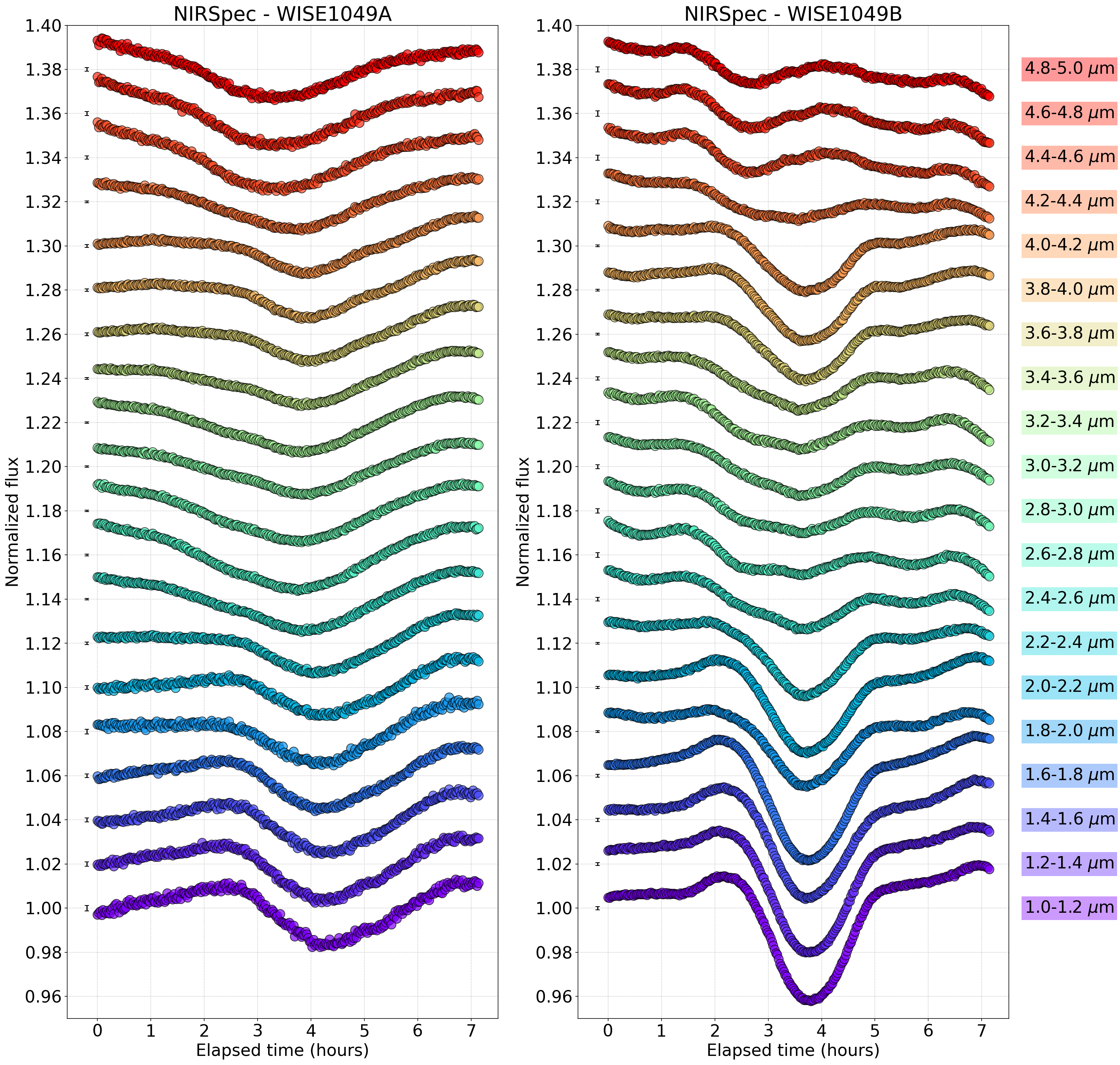}
\caption{NIRSpec light curves for WISE 1049A (left) and WISE 1049B (right), with 0.2 $\mu$m wavelength bins and a cadence of 90 s. Light curves have been normalized to their median values to highlight fractional variations. The light curves are offset with constants to visually separate them on the plot. The uncertainty for each light curve is shown as an error bar on the left side of the plot.}
\label{fig:lc_nrs}
\end{figure*}

\subsection{Variability map and light curves}
\label{sec:varmap}

We present variability maps of the MIRI and NIRSpec observations for WISE 1049A and B in Fig.  \ref{fig:varmap_miri} and \ref{fig:varmap_nrs}. The variability map is a 2D representation of the flux (normalized with the median flux at each time step) as a function of wavelength and time. 
The spectral resolution across the MIRI bandpass varies from R $\sim$ 40 at 5 $\mu$m to R $\sim$ 160 at 10 $\mu$m, and the spectral resolution across the NIRSpec bandpass varies from R $\sim$ 30 at 1 $\mu$m to R $\sim$ 300 at 5 $\mu$m. To account for this changing wavelength bin width, we interpolate the spectral TSO array onto a uniformly spaced wavelength grid with a spectral bin width of 0.01 $\mu$m. We then divide each row (i.e. the light curve at each wavelength point) by the median flux of that row, in order to get the fractional variation around the median. We then plotted the spectral light curve array as a colormap in time and wavelength, with the color bar showing the normalized fluxes from 0.98 to 1.02.

The MIRI variability maps of both WISE 1049A and B show a change of behaviour around 8.5 $\mu$m, which corresponds to the beginning of the silicate absorption feature, displaying almost anti-correlated brightness variations for shorter and longer wavelengths. A narrow stripe of distinct light curve behaviour is also seen around the 6.5 $\mu$m water absorption feature for both components. The variability map around 5--9 $\mu$m displays a steep drop of brightness in the first 15--20 min of the observation; this is the downward ramp effect reported by \citep{Bell2023} and also seen in \citep{Biller2024}. This effect appears to be negligible after the first half-hour of the observation. Therefore in the following analysis we removed the MIRI data points in the first 30 minutes of the observation.

The NIRSpec variability maps feature prominent troughs of brightness for both components. For WISE 1049A, the minimum at 1 $\mu$m appears around 4.5 hours and shifts gradually to around 3.5 hours at 5 $\mu$m. The variability map for WISE 1049B shows some regions of very abrupt changes, and narrow strips of distinct behaviour are seen around the 3.3 $\mu$m CH$_4$ feature and several water and CO bands. 
Both maps show distinct regions of different light curve behaviour as a function of wavelength, with clear breaks occurring around 2.5 and 4.3 $\mu$m for both binary components, similar to the findings from \cite{Biller2024}.

\subsubsection{MIRI light curves}

\begin{figure*}
\centering
\begin{minipage}{.66\textwidth}
  \centering
  \includegraphics[width=1\linewidth]{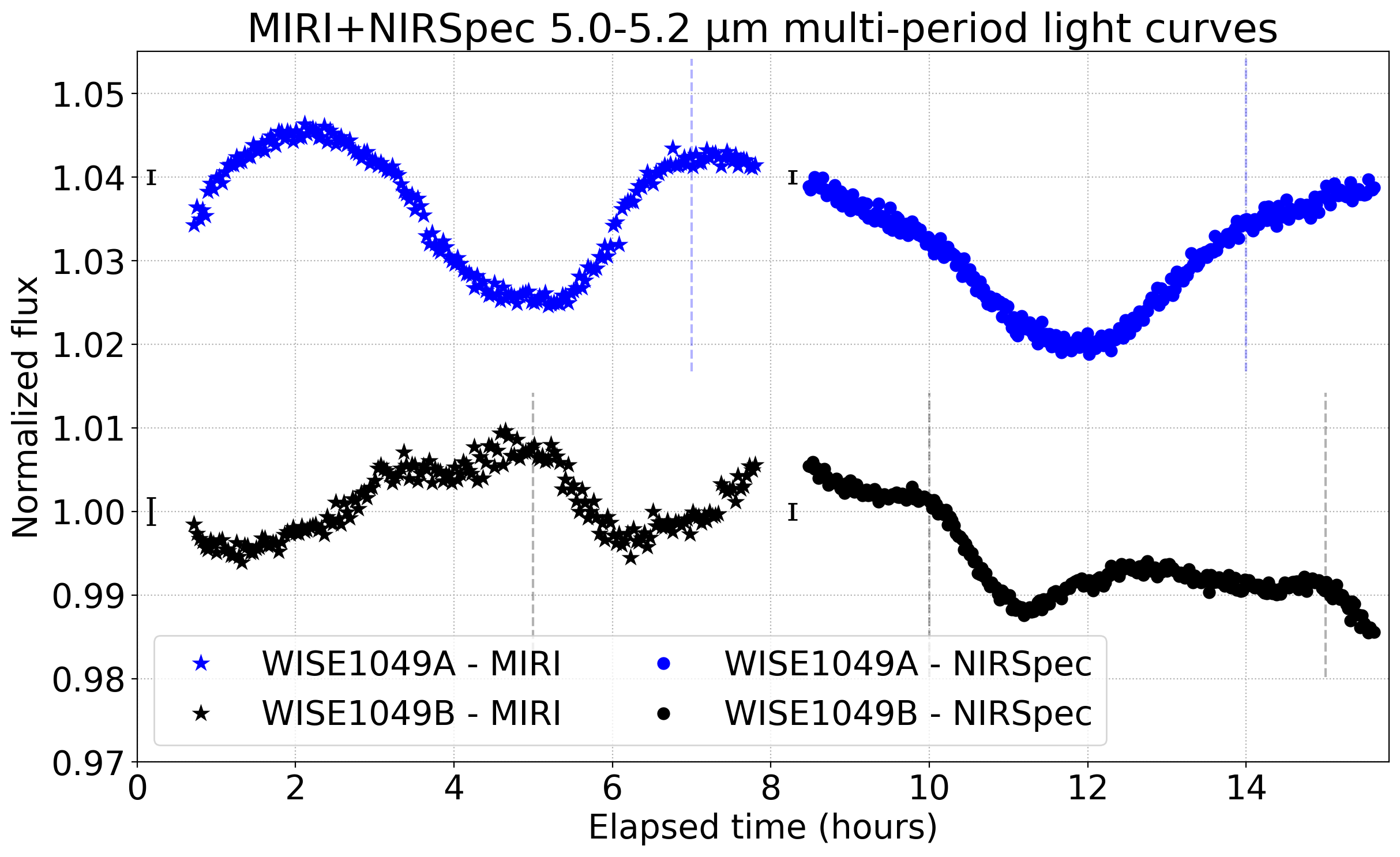}
\end{minipage}
\begin{minipage}{.30\textwidth}
  \centering
  \includegraphics[width=0.9\linewidth]{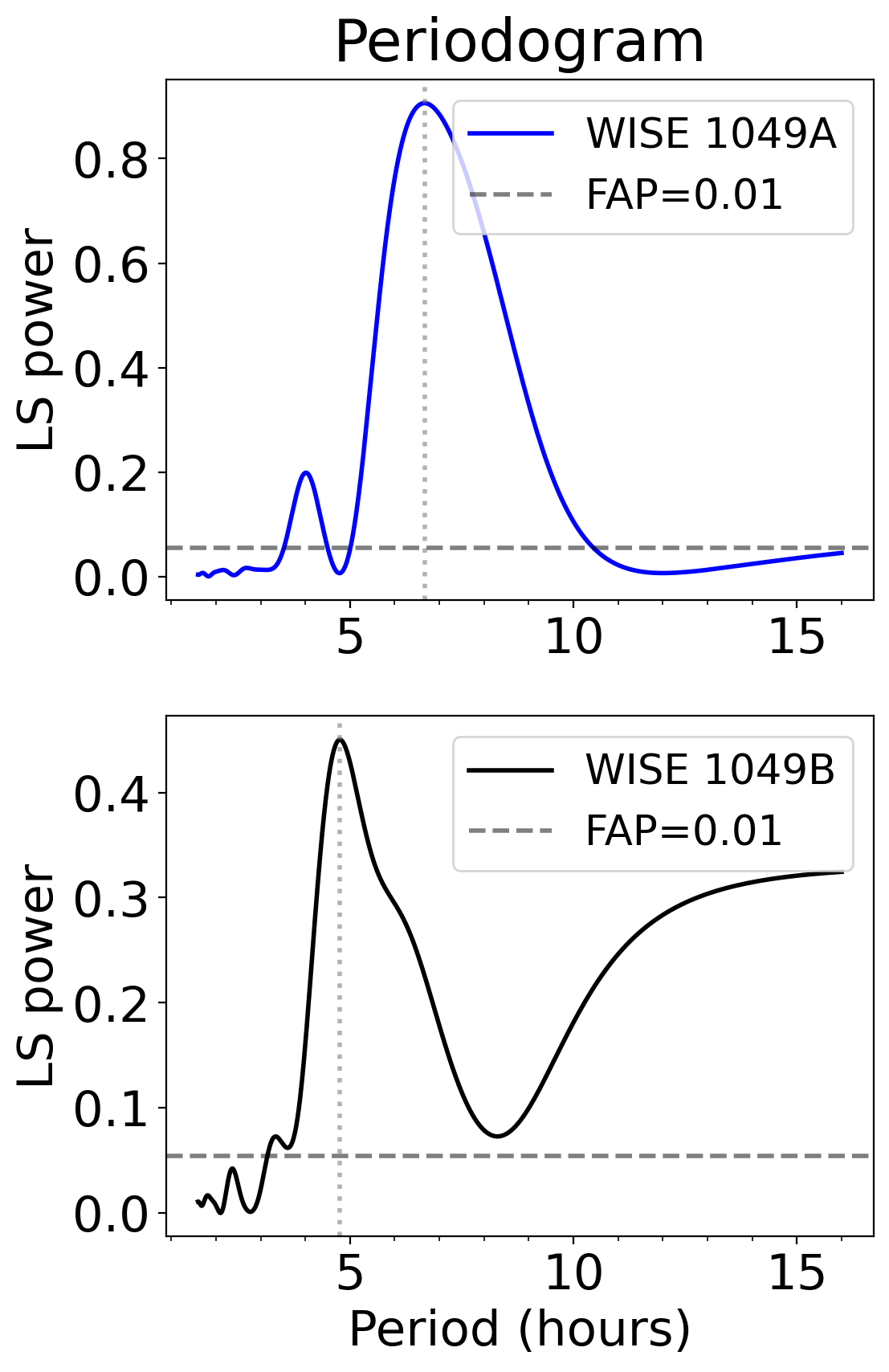}
\end{minipage}
\caption{Left: Combined light curves over $\sim$15 hours of WISE 1049 A and B from the overlapping wavelengths regions covered by both MIRI and NIRSpec (5--5.2 $\mu$m). The MIRI and NIRSpec light curves have both been normalized separately by their respective median values and offset with a constant; there remains an unknown scaling factor between the two sets of observations. Error bars are estimated with $\sigma_\mathrm{pt}$ \citep{Biller2024} and are shown on the left of each light curve. Vertical dashed lines are drawn on the light curves of WISE 1049A and B to indicate their respective rotational period of 7 h and 5 h. Right: Lomb–Scargle periodograms for WISE 1049A (upper panel) and WISE 1049B (lower panel), considering periods up to 16 h, and using the MIRI + NIRSpec light curves shown on the left panel. The horizontal black dashed line shows the 99 per cent false alarm probability power levels, and the vertival black dotted lines mark the periods where maximum LS power is found.}
\label{fig:combined_lc}
\end{figure*}

From the spectral TSO arrays from MIRI and NIRSpec observations, we construct binned light curves for WISE 1049A and B in Fig. \ref{fig:lc_miri} and Fig. \ref{fig:lc_nrs}. The MIRI observations have been binned by a factor of 100 in time, for a resulting exposure time of 129 s for each data point shown. Each light curve was constructed by integrating the flux over a wavelength bin of 0.5 $\mu$m, taking into account the changing spectral resolution across the MIRI bandpass. 
The light curves for each wavelength bin were then normalized by their median values to highlight fractional variations at each wavelength.
Uncertainties for each light curve were estimated using the same $\sigma_\mathrm{pt}$ method as in \cite{Biller2024}, i.e. the standard deviation of the light curve subtracted from a version of itself shifted by one step in time, divided by $\sqrt{2}$. 
The first 30 min of MIRI light curves were removed due to the steep downward ramp effect seen in the variability maps in Fig. \ref{fig:varmap_miri}. 

Light curves of both WISE 1049A and B in Fig. \ref{fig:lc_miri} show variability at all wavelengths from 5 $\mu$m to $\sim$14 $\mu$m, with WISE 1049A displaying higher variability than B in these longer wavelengths in this specific epoch. The light curve shapes vary as a function of wavelength, pointing to different mechanisms affecting different wavelengths. The light curves show distinctive shapes below vs. beyond 8.5 $\mu$m.
For WISE 1049A, the light curves below 8.5~$\mu$m exhibit a minimum around 5 hours, whereas the light curves from 8.5--10~$\mu$m show a maximum at the same time. For WISE 1049B, a minimum is seen around 1 hour for light curves below 8.5 $\mu$m, while for light curves from 8.5--10 $\mu$m a local maximum is found at this time. A similar trend is seen around 5 hours, where light curves below 8.5 $\mu$m show a local maximum while 8.5--10 $\mu$m show a local minimum. In general, the MIRI light curves show anti-correlation below vs. beyond 8.5 $\mu$m, as seen earlier in the variability maps.

\subsubsection{NIRSpec light curves}

From the spectral TSO arrays from NIRSpec observations, we construct binned light curves for WISE 1049A and B in Fig. \ref{fig:lc_nrs} in the same way as for MIRI. 
The NIRSpec observations have been binned by a factor of 200 in time, for a resulting exposure time of 90 s for each point shown. For each light curve, we integrated the flux over a wavelength bin of 0.2 $\mu$m.

In the NIRSpec wavelengths, both WISE 1049A and B show $>$1\% variability, with the B component more variable than A. 
WISE 1049B light curves show a dramatic minimum around 4 hours into observation for 1--2.6 $\mu$m and 3.6--4.2 $\mu$m. The light curves from 4.4--5 $\mu$m shows an anti-correlation from the light curves below 4.4 $\mu$m. 
For WISE 1049A, the minimum at 1 $\mu$m appears around 4.5 hours and shifts gradually to around 3.5 hours at 5 $\mu$m. Since the NIRSpec light curves are cross-contaminated from the tile event at a level of 3\%, combined with the fact that the variability of A component is only around 1 \% and $\sim$4 times smaller than that of B in the shortest wavelengths, this means that the minima seen in the WISE 1049A light curves could contain signals from WISE 1049B.

\subsection{MIRI + NIRSpec multi-period light curves and periodogram analysis}

Since both the MIRI and the NIRSpec observations cover the wavelength range of 5--5.2 $\mu$m, we can obtain light curves in this window over 15 hours for the binary, which covers $\sim$2 rotations of A and $\sim$3 rotations of B. These multi-period lightcurves are plotted in the left panel of Fig. \ref{fig:combined_lc}. We found that the shapes of the light curves for both components differ significantly from period to period and this suggests rapid short-term evolution of these brown dwarfs.

We used the Lomb–Scargle periodogram \citep{VanderPlas2018} as implemented in \textsc{astropy} to determine the periodicities in our light curves. 
Periodogram results are shown in the right panel of Fig. \ref{fig:combined_lc}. We considered periods from 1.2 to 16 h and calculated the 99 per cent false alarm probability power using the built-in bootstrap method. We found a peak power in the periodogram around 6.7 hours for WISE 1049A and close to 5 hours for WISE 1049B, which are consistent with previous period measurements of $\sim$7 h for A and $\sim$5 h for B (e.g., \citealt{Apai2021, Fuda2024}).

\subsection{Comparison with previous epoch}
\label{sec:bol}

\begin{figure*}
\centering
\begin{minipage}{.6\textwidth}
  \centering
  \includegraphics[width=0.98\linewidth]{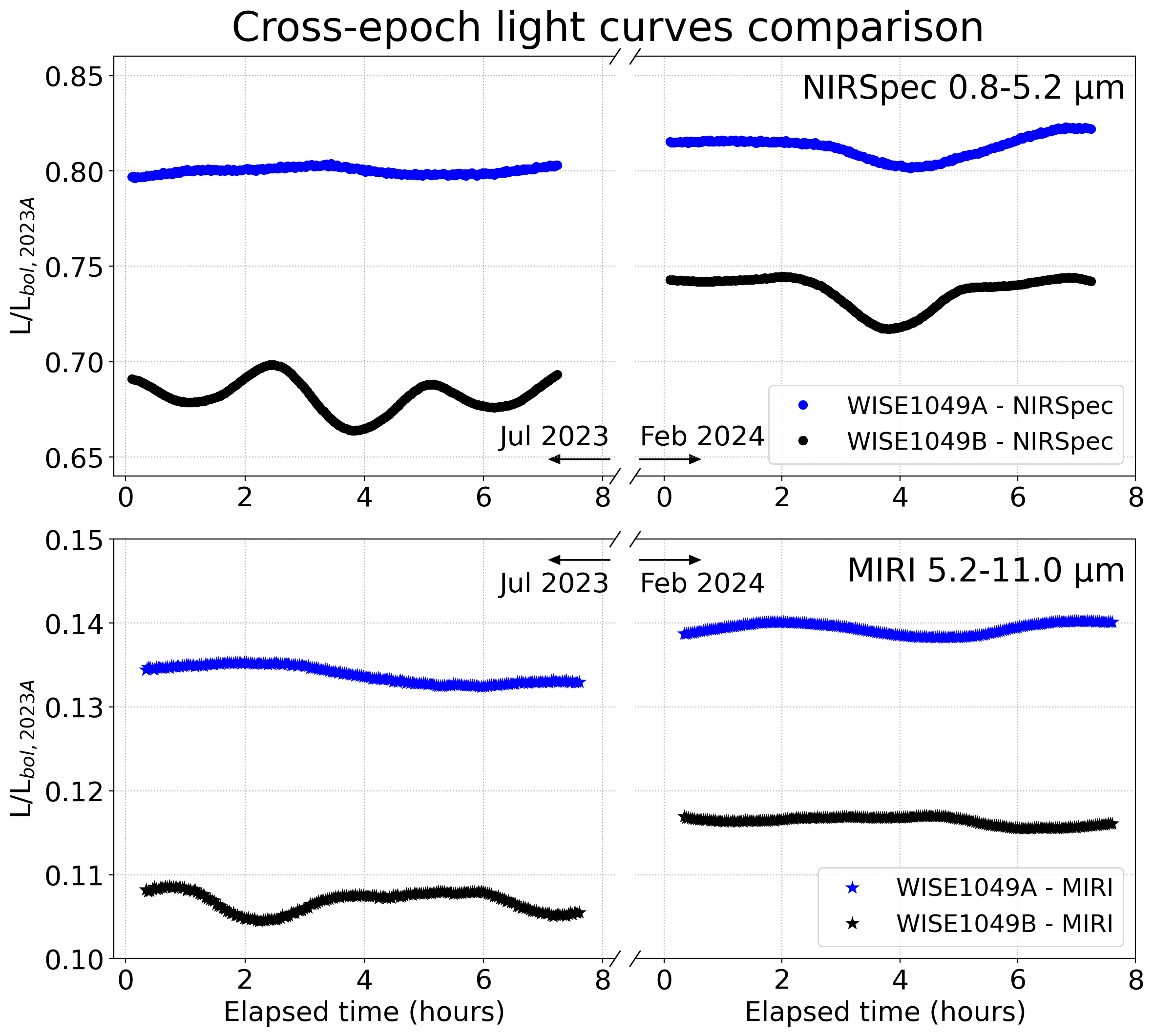}
\end{minipage}
\begin{minipage}{.38\textwidth}
  \centering
  \includegraphics[width=0.95\linewidth]{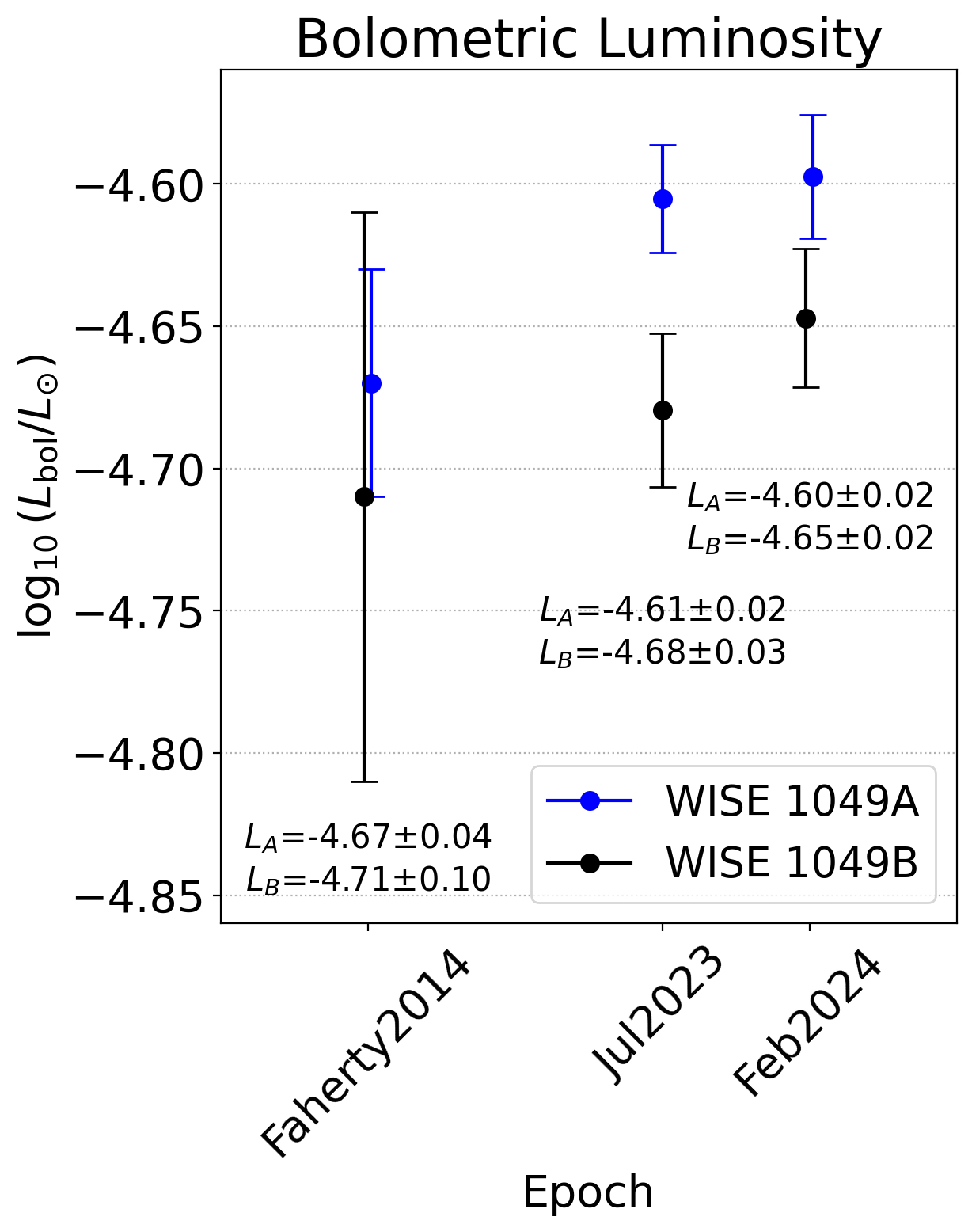}
\end{minipage}
\caption{Left: Cross-epoch flux comparison of WISE 1049 A and B over the full NIRSpec and MIRI bandpass between the July 2023 epoch \citep{Biller2024} and the Feb 2024 epoch (in this paper). 
Right: Estimated bolometric luminosity of WISE 1049AB from this epoch, compared with values reported in the literature and previous epoch (\citealt{Faherty2014, Biller2024}). We do not claim a significant cross-epoch brightness variation in the bolometric luminosity as the measured $L_\mathrm{bol}$ between July 2023 and Feb 2024 epochs agree within error bars.}
\label{fig:bol}
\end{figure*}

We constructed absolute-flux, broadband "white-light" light curves across the full NIRSpec and MIRI wavelengths and compared them with the previous epoch in July 2023 as reported in \cite{Biller2024}. For NIRSpec, we summed the flux from 0.8--5.2 $\mu$m. For MIRI, we summed the flux from 5.2--11 $\mu$m. We converted the flux into units of W/m$^2$, adopting a distance of 1.9960$\pm$0.0002 pc to WISE 1049AB \citep{Bedin2024}, and then expressed all light curves in fractions of the estimated bolometric luminosity of WISE 1049A from the July 2023 epoch (taking $L_\mathrm{bolA, 2023}/L_\odot$=-4.61). 
The left panel of Fig. \ref{fig:bol} shows the resulting white light curves from both epochs side by side on the same scale. 
 Looking at the broadband NIRSpec and MIRI light curves on the left panel of Fig. \ref{fig:bol}, we see some cross-epoch variations within the NIRSpec and MIRI bandpasses. 
Both the NIRSpec and MIRI broadband fluxes for the two components increased slightly compared to the 2023 epoch.

We then estimated the bolometric luminosity as in \cite{Biller2024}, taking into account both the intrinsic variability of the objects and the range of fractions of their full luminosity covered in the {\sl JWST} bandpass, $L_\mathrm{JWST}$. We added the luminosity from NIRSpec and MIRI together to be $L_\mathrm{JWST}$, and then assumed that $L_\mathrm{JWST}/L_\mathrm{bol}$ ranges from 0.9 to 0.97. This estimation is based on calculated $L_\mathrm{JWST}/L_\mathrm{bol}$ from \cite{Biller2024} using ExoRem model grids \citep{Charnay2018} ranging from 1100 to 1900 K and various $\log g$. Then, the final range of estimated bolometric luminosity is given by $\frac{L_{\mathrm{min,JWST}}}{0.97}$ to $\frac{L_{\mathrm{max,JWST}}}{0.9}$. 
This results in a bolometric luminosity of $L_\mathrm{bol}/L_\odot$=-4.60$\pm$0.02 for WISE 1049A, and $L_\mathrm{bol}/L_\odot$=-4.65$\pm$0.02 for WISE 1049B.
We compared the measurements from this epoch to the July 2023 epoch and previously reported values in \cite{Faherty2014} in the right-hand panel of Fig. \ref{fig:bol}.
All three of the $L_\mathrm{bol}$ measurements for WISE 1049B agree within error bars, and the measured $L_\mathrm{bol}$ for WISE 1049A from the July 2023 and Feb 2024 epoch agreed within error bars. Therefore, we do not claim a significant cross-epoch brightness variation in the bolometric luminosity for both components of the binary. 

Errors in our estimation of bolometric luminosity come from multiple sources, including: 1) varying $F_\mathrm{NIRSpec}/F_\mathrm{bol}$ and $F_\mathrm{MIRI}/F_\mathrm{bol}$, i.e. the fraction of flux covered by the NIRSpec and MIRI bandpasses respectively; 2)
blending of the A and B components due to the closer separation in 2024 compared to 2023; 3) cross-contamination of the A and B components caused by the mirror tilt event; and 4) changes in {\sl JWST} calibration from 2023 to 2024. The blending between components should cause the fluxes of both components to be closer to each other, as is observed in this epoch. The level of this contamination should be $<$5\%.

\section{Analysis and discussion}
\label{sec:analysis}

\subsection{Light curve clusters}

\begin{figure*}
\centering
\begin{minipage}{0.492\textwidth}
  \centering
  \includegraphics[width=1\linewidth]{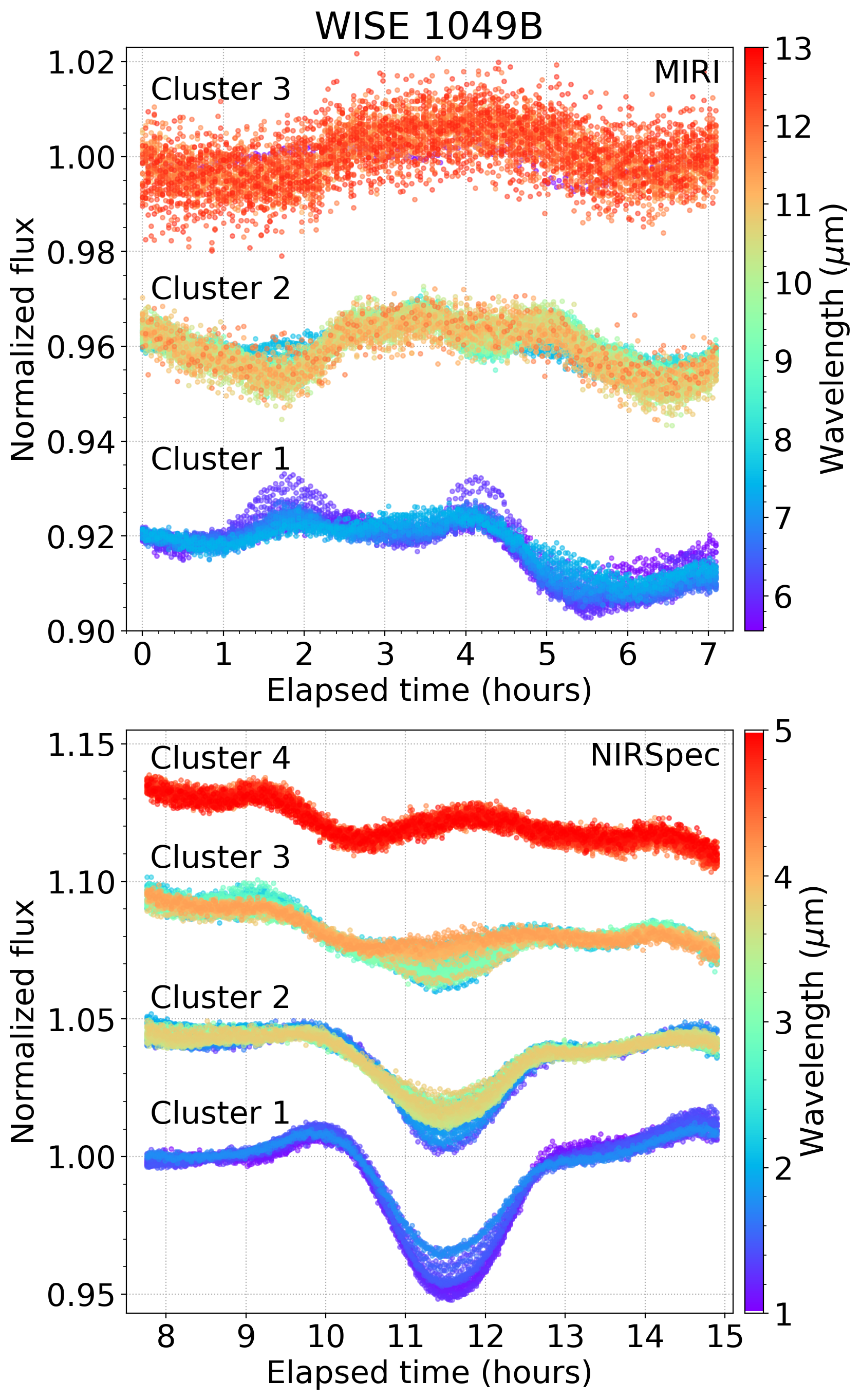}
\end{minipage}
\vspace{5pt}
\begin{minipage}{0.492\textwidth}
  \centering
  \includegraphics[width=1\linewidth]{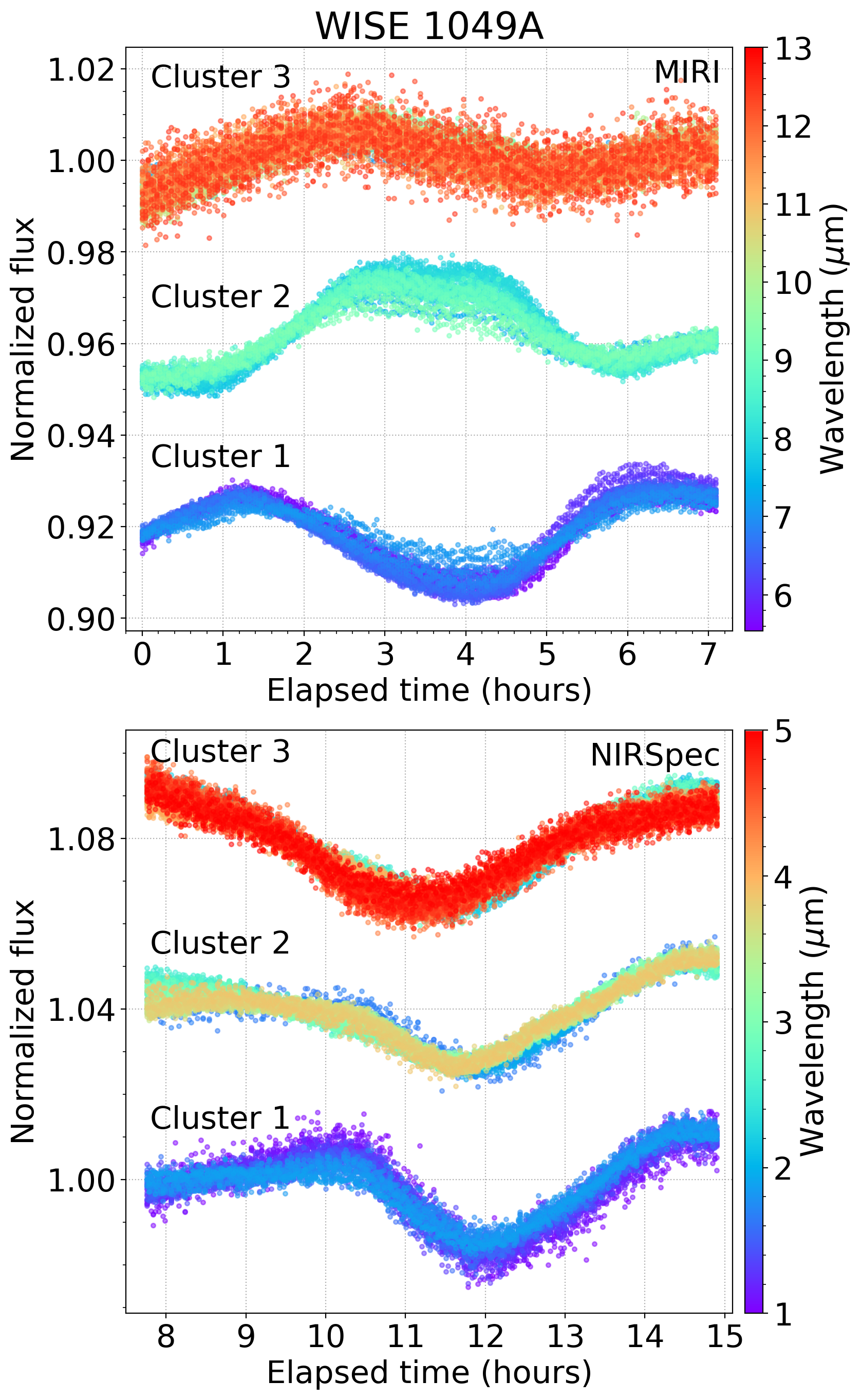}
\end{minipage}
\caption{Light curve clusters from MIRI and NIRSpec assigned by the k-means clustering algorithm for WISE 1049B (left) and WISE 1049A (right).}
\label{fig:clusters}
\end{figure*}

\begin{figure*}
\centering
\begin{minipage}{1\textwidth}
  \centering
  \includegraphics[width=1\linewidth]{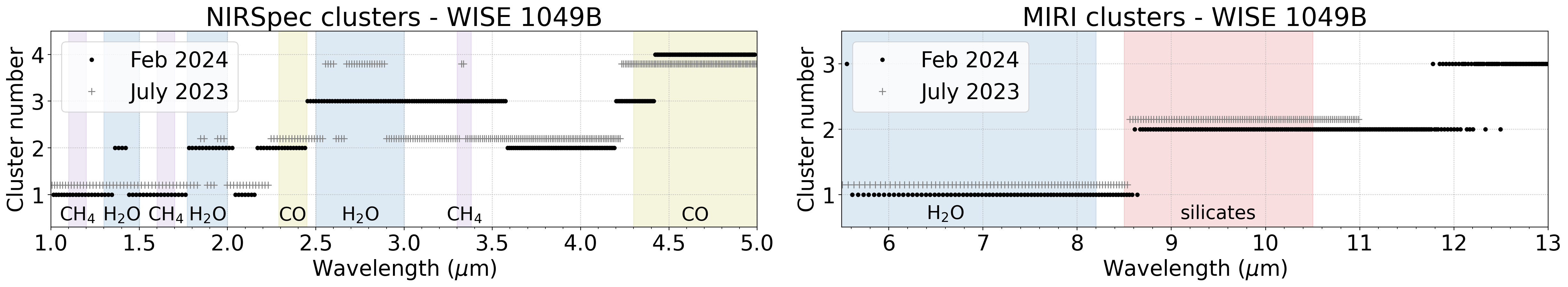}
\end{minipage}
\vspace{5pt}
\begin{minipage}{1\textwidth}
  \centering
  \includegraphics[width=1\linewidth]{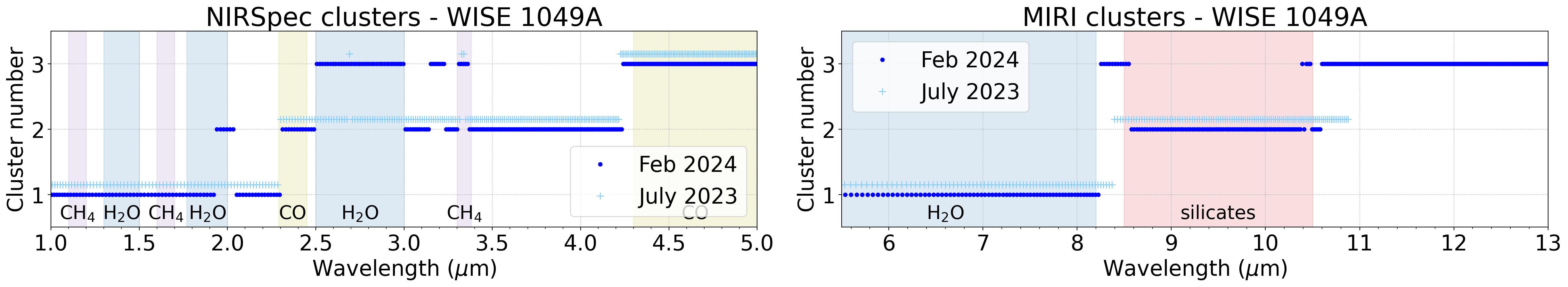}
\end{minipage}
\caption{The assigned cluster numbers as a function of wavelength for WISE 1049B (upper panels) and WISE 1049A (lower panels). The cluster assignments from the July 2023 epoch \citep{Biller2024} are shown in lighter colours on the side for comparison. Various molecular absorption bands are labelled with shaded areas. The wavelengths at which clusters transition remained broadly consistent over the two epochs observed.}
\label{fig:clusterbreaks}
\end{figure*}

\begin{figure*}
\includegraphics[width=1\textwidth]{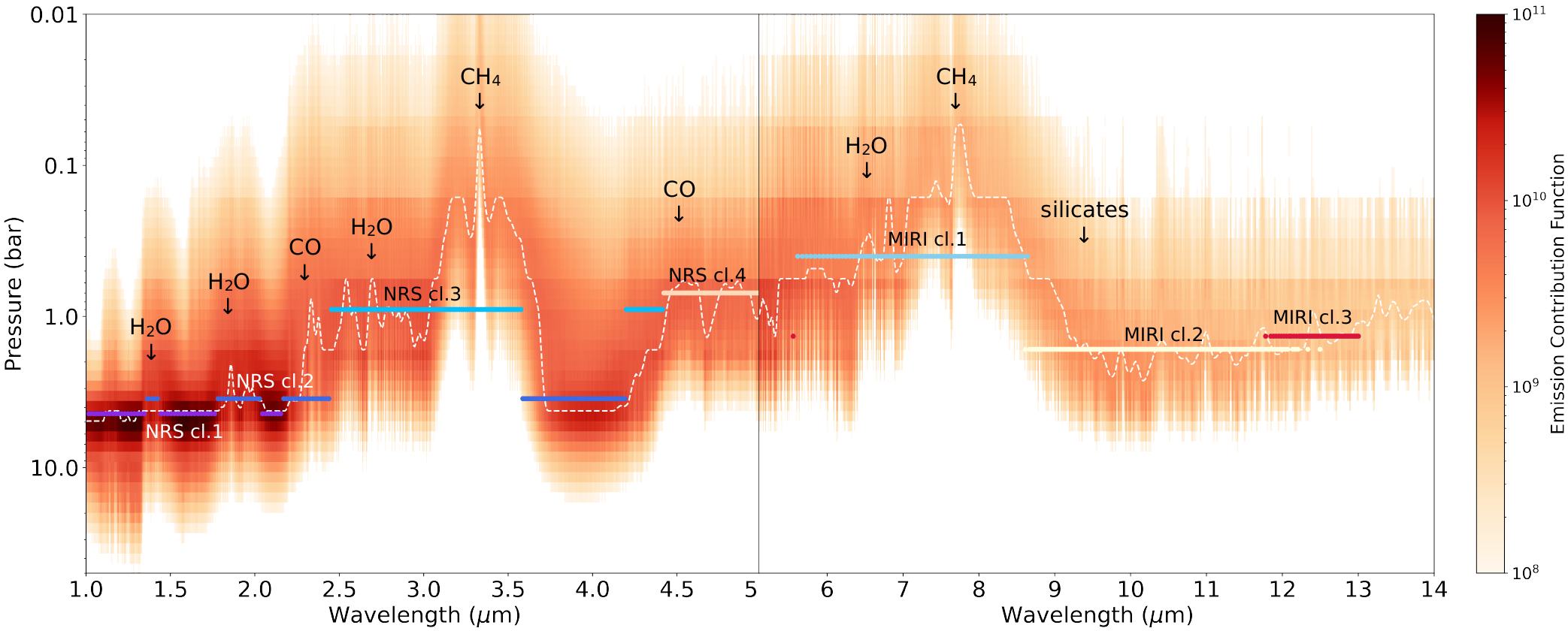}
\caption{Contribution function from 1--14 $\mu$m calculated from a Sonora Diamondback model with $T_\mathrm{eff}$=1300K, logg=5, $f_\mathrm{sed}$=8, and solar metallicity. The assigned clusters of WISE 1049B from NIRSpec and MIRI are overlaid on the contribution function at the typical pressure levels within each cluster. The white dotted line shows the pressure where the maximum flux is emitted at each wavelength. Transitions in light curve behaviour are found to align with changes in the pressure levels.}
\label{fig:cf}
\end{figure*}

The narrowband light curves presented in the previous sections showed that the light curves at certain wavelengths share similar patterns. This suggests that the variability observed at these specific wavelength regions are likely driven by the same underlying mechanisms \citep{Biller2024, McCarthy2024}. Motivated by this, we group light curves with similar shapes using a clustering algorithm.  Each cluster then corresponds to a set of wavelengths that can be mapped to distinct pressure levels within the atmosphere. This allows us to link the observed variability to specific vertical layers in the atmosphere and disentangle the contributions from different variability-driving processes.

Following the procedure in \cite{Biller2024}, we use a K-means clustering algorithm from \texttt{scikit-learn} to group the MIRI and NIRSpec light curves by shape. The input data consists of the array of normalized single-wavelength-bin light curves at the original MIRI/NIRSpec spectral resolution (i.e. the variability maps displayed in Fig. \ref{fig:varmap_miri} and \ref{fig:varmap_nrs}) and their corresponding wavelengths. The K-means algorithm then determines the position of cluster centroids and assigns each light curve to a cluster. We ran 10 tests with $n_\mathrm{cluster}$ from 1 to 11 and determined the optimal number of clusters using the elbow point method implemented in \texttt{KneeLocator} from \texttt{kneed} \citep{Satopaa2011}. The final output is the assigned cluster number for each wavelength point.

The resulting clusters of light curves for both binary components are plotted in Fig. \ref{fig:clusters}, and the cluster assignments as a function of wavelength are shown in Fig. \ref{fig:clusterbreaks}. 
For WISE 1049B, we found 4 clusters in the NIRSpec data and 3 in MIRI. NIRSpec Cluster 1 and 2 cover the shortest wavelengths and capture the pronounced deep trough pattern at around 3.8~h. Cluster 4 represents a dramatically different behaviour with a smaller variability amplitude at the red end of the NIRSpec wavelength range, while Cluster 3 shows behaviours in transition. For MIRI, Cluster 1 includes light curves below 8.5 $\mu$m, while Clusters 2 and 3, which have similar shapes and are in phase, cover wavelengths above 8.5 $\mu$m. Specifically, Cluster 2 spans 8.5 to $\sim$12~$\mu$m, and Cluster 3 includes wavelengths beyond 12 $\mu$m.

For WISE 1049A, we identified 3 clusters in NIRSpec and 3 in MIRI. In NIRSpec, Cluster 1 corresponds to the shortest wavelengths, similar to Cluster 1 for WISE 1049B. Cluster 2 spans intermediate wavelengths, which mostly corresponds to the 2nd cluster for WISE 1049B. Cluster 3 covers the H$_2$O band from 2.5--3.0 $\mu$m and the CO fundamental band from 4.3 $\mu$m onward, roughly corresponding to Clusters 3 and 4 for WISE 1049B.
In the MIRI data, Cluster 1 includes the shorter wavelengths (below 8.5 $\mu$m) with a trough centered at 3--5 hours. Cluster 2 spans 8.5--10.5 $\mu$m, with light curves almost anti-correlated with Cluster 1. Cluster 3 covers wavelengths beyond $\sim$10.5 $\mu$m with light curves slightly phase-shifted from Cluster 2. Interestingly, the transition between Cluster 2 and 3 for WISE 1049A happens at a shorter wavelength than that for WISE 1049B, making Cluster 2 for WISE 1049A just covering the silicate absorption band between 8.5--10.5 $\mu$m. This difference in cluster assignment around the silicate feature between WISE 1049A and B potentially reflects their different spectral types and evolutionary stages: the A component, a late L-type, shows signs of cloud opacity in the silicate feature, whereas this feature is less evident in the early T-type B component. This is also consistent with the weak detection of this silicate feature in WISE 1049A's spectra, but not in B's, as shown earlier in Section \ref{sec:results}.

The cluster assignments as a function of wavelength are compared to the assignments in the July 2023 epoch in Fig. \ref{fig:clusterbreaks}. We found that the transition between clusters happened at similar wavelengths: around 2.3--2.5 $\mu$m, corresponding to a CO absorption bandhead; around 4.3 $\mu$m, marking the start of the CO fundamental band; and at 8.5 $\mu$m, where the water absorption band ends and the silicate feature begins.
Although we found a total of 4 clusters from NIRSpec instead of 3 in the previous epoch for WISE 1049B, the difference is mainly in the wavelength range of 2.5--3.5 $\mu$m: While the previous epoch assigned light curves in this range to either Cluster 2 or 3, our clustering put light curves in these wavelengths into an additional new cluster which shows mixed behaviours from its adjacent clusters.

Comparing the cluster assignments with the previous epoch, the broad consistency of these transition points suggests that, despite the somewhat chaotic evolution of the light curves, the underlying mechanisms driving the variability likely remained the same over the two observed epochs.

\subsection{Contribution from different pressure levels}
\label{sec:cf}

To identify the pressure levels probed by each light curve cluster, we rely on comparison with the thermal emission contribution computed from models. The contribution function quantifies the flux emitted by each pressure layer at specific wavelengths, defined as follows \citep{Lothringer2018} \footnote{Note that the original equation in the paper is missing a minus sign.}:

\begin{equation}
\mathrm{CF}(p, \lambda)=B_\lambda * e^{-\tau_{p, \lambda}} \frac{d \tau_{p, \lambda}}{d p}
\end{equation}

where $B_\lambda (T)$ is the blackbody radiation at a given pressure-temperature level, and $\tau_{p, \lambda}$ is the optical depth at pressure $p$ and wavelength $\lambda$.
Using the pressure-temperature profiles from the Sonora Diamondback models \citep{Morley2024} and the \textsc{picaso} package \citep{Batalha2019}, we computed an contribution function for WISE1049B, shown in Fig.~\ref{fig:cf}. We used a cloudy model in chemical equilibrium with $T_\mathrm{eff}$=1300 K, $\log g$=5 (cgs), $f_\mathrm{sed}$=8 and solar metallicity. This model reasonably matches the effective temperature and $\log g$ derived from evolutionary model fits to the bolometric luminosity of both components, as calculated in Section. \ref{sec:bol}. Therefore, this contribution function serves as a indicative guide for interpreting how top-of-atmosphere (TOA) structures at different pressure levels influence the observed light curves across wavelengths.

By overlaying the WISE 1049B cluster assignments onto the contribution function, we found that transitions in light curve behaviour often align with changes in the pressure levels. We identified roughly 3 distinct pressure levels corresponding to the light curve clusters: 1) A deep pressure level around 2--10 bar that probes the hottest part of the atmosphere and exhibiting the largest variability amplitudes. This is reflected in the light curves of NIRSpec clusters 1 and 2 where deep troughs are seen.  
2) A high-altitude level $<\sim$ 1 bar, probed by NIRSpec Cluster 4 + MIRI Cluster 1 between 4.3 --8.5 $\mu$m, and by NIRSpec Cluster 3 between 2.5 --3.6 $\mu$m. 
These wavelengths align with various H$_2$O and CH$_4$ absorption bands, as well as the CO fundamental band between 4.3--5.0 $\mu$m.
3) An intermediate pressure level at $\sim$1--3 bar probed by MIRI Cluster 2 \& 3 at wavelengths $>$8.5 $\mu$m. This level begins at wavelengths coinciding with the small-grain silicate absorption features between 8.5 and 11 $\mu$m. \citep{Cushing2006, Suarez2022}. 
The pressure levels found in this epoch of data broadly agree with the findings from the previous epoch in \cite{Biller2024}. In that study, high pressure levels deep in the atmosphere were found to drive double-peaked variability at wavelengths $<$2.3 $\mu$m and $>$8.5 $\mu$m, whereas lower pressure, high-altitude levels produced light curve behaviours between 4.2 and 8.5 $\mu$m.

\subsection{Deviation spectra as an indicator of variability mechanism}
\label{sec:devspectra}



\begin{figure}
\includegraphics[width=1\columnwidth]{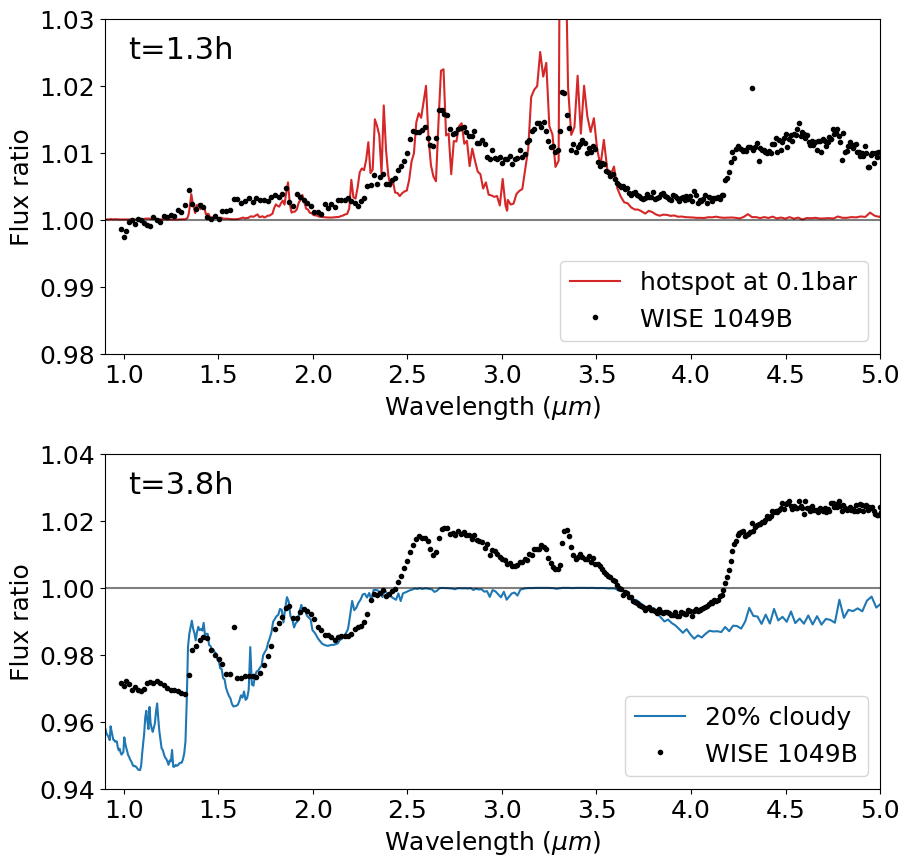}
\caption{Snapshots of the deviation-from-mean spectra (black dots) at 1.3 h and 3.8 h into the NIRSpec observation compared to models featuring hot spots and clouds from \citet{Morley2014}. The hot spot model (red line) represents an atmosphere with injected energy at 0.1 bar covering 0.15\% of the atmosphere. The cloud model (blue line) represents an atmosphere with 20\% cloud coverage.}
\label{fig:dev_spec}
\end{figure}

The time-resolved deviation-from-mean spectra provide insights into the mechanisms driving variability at different time points in the rotation (e.g., \citealt{McCarthy2024}). 
We compute the deviation spectrum of WISE 1049B at each time point (with 90 s cadence for NIRSpec and 129 s cadence for MIRI) by taking the ratio of the spectrum at that time point to the mean spectrum over the entire 7–8 hours of NIRSpec or MIRI observations. By this definition, a deviation spectrum represents the percentage difference of an observed spectrum at a given time point compared to the time-averaged spectrum.

We then compare the time-resolved deviation spectra with predictions from models featuring two different variability-driving mechanisms, i.e. hot spots and clouds, from \cite{Morley2014}. In these models, we have a baseline model of a brown dwarf at $T_\mathrm{eff}$=1000 K and $\log g$=5 with clear atmosphere in chemical equilibrium. The hot spot model is produced by injecting energy at 0.1 bar to the baseline model and could represent any heating mechanism. The deviation spectrum predicted by the hot spot model is then calculated by taking a fraction of the hot spot spectrum to represent a portion of the atmosphere covered by hot spots, then dividing by the baseline spectrum. The cloudy model is produced by including cloud species with $f_\mathrm{sed}$=5 and the predicted deviation spectrum is calculated by taking a fraction of the completely cloudy spectrum to represent partial cloud coverage, and then dividing by the baseline model. 
These models represent the percentage difference between a spectrum with variability features and the spectrum of a uniform, non-variable atmosphere (the baseline model). By comparing the observed deviation spectra to the models, we assume that the time-averaged spectrum is representative of a featureless, uniform atmosphere, and that any deviations from the time-average are caused by the variability features. In this section we mainly focus on WISE 1049B, as its NIRSpec light curves are less affected by the mirror tilt event and exhibit clearer variability features.

Snapshots of the observed deviation spectrum at two selected time points are plotted in Fig. \ref{fig:dev_spec}, along with predictions from hot spot and cloudy models. 
For WISE 1049B, the deviation spectrum at 1.3 h aligns well with a hot spot model with 0.15\% spot coverage, assuming only the total area of hot spots without specifying their shape or the total number of spots. Both predicted and observed deviation spectrum show excess flux between 2.3--3.6 $\mu$m, where various molecular absorption band lies (i.e., CO at 2.3--2.5 $\mu$m, H$_2$O at 2.5--3 $\mu$m and CH$_4$ at 3.3--3.4 $\mu$m). Notably, light curves between 2.3–3.6 $\mu$m in Fig. \ref{fig:lc_nrs} show a peak at this exact time, suggesting that these peaks are likely driven by hot spots associated with these molecules. \cite{Apai2017} showed that an elliptical bright spot can introduce a localized bump in rotational light curves. The overall irregular shape of the 2.3–3.5 $\mu$m light curves could be the result of multiple hot spots in combination with other possible sources of variability. While the hot spot models only represent an atmosphere with regions of elevated temperature, one possible physical mechanism for forming such hot spots is vertical mixing through convective columns, where hotter air from deeper in the atmosphere is transported up to the higher altitudes probed by these wavelengths \citep{Tremblin2020}.

On the other hand, at 3.8 h, the deviation spectrum matches well a model with 20\% cloud coverage, especially in the 1--2.5 $\mu$m and 3.5--4.0 $\mu$m ranges. This timing coincides with the occurrence of dramatic troughs observed in the 1--2 $\mu$m light curves in Fig. \ref{fig:lc_nrs} at 3.8 h, providing evidence that the light curve patterns at these wavelengths are shaped by patchy clouds rotating into view. 
Interestingly, neither model provides a good match to the spectra from 4.3--5 $\mu$m, suggesting that they are missing some variability mechanisms at these wavelengths, most likely related to carbon chemistry, as this range is dominated by the CO fundamental band (See also discussion in Section \ref{sec:gcm}).

Note that the two time points discussed here were selected because at these two times the spectrum contains almost entirely contribution from one of the two different mechanisms, which is good for demonstration purposes. At a more arbitrary time point, the spectrum typically shows a blend of both mechanisms.
It is also important to remember that these comparisons rely on 1D models with simplified assumptions. While they provide an indicative guide for interpreting the spectra, they do not fully capture the complexity of real atmospheric processes. As such, the connection between light curve/spectrum behaviour and the variability-driving mechanisms discussed in this section should be considered a tentative hypothesis rather than a definite statement. Also, it is likely that multiple mechanisms are at play simultaneously.

\subsection{Comparing both epochs with a general circulation model} 
\label{sec:gcm}

\begin{figure*}
\includegraphics[width=1\textwidth]{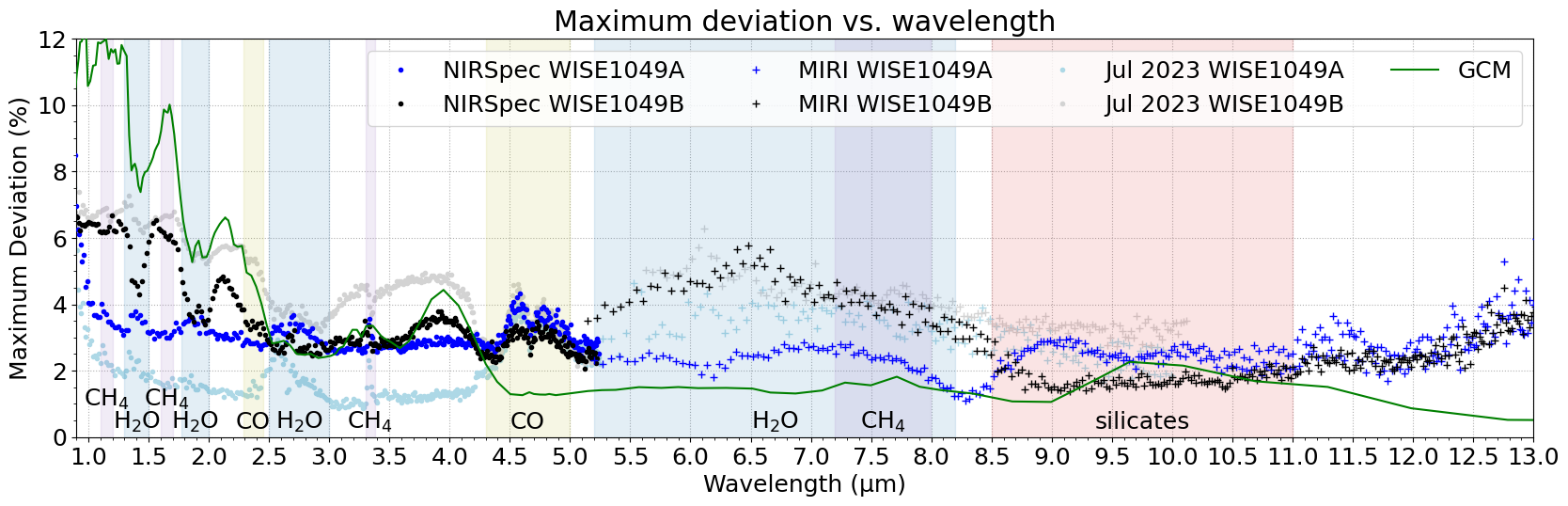}
\caption{Maximum deviation as a function of wavelength. Prediction from a general circulation model (GCM) with parameters adapted to those of WISE 1049B (gravity of
588 ms$^{-2}$, radius of 1.05 $M_\mathrm{Jup}$ and rotation period of 5.28 h) is shown as a green line. The absorption bands of various molecules are marked with shaded areas. GCM predicts increased variability in the continuum between absorption bands. Peaks of variability are observed in the continuum between water bands below 2.5 $\mu$m, and at the 2.7 $\mu$m H$_2$O, 3.3 $\mu$m CH$_4$, 4.5 $\mu$m CO, and 6.5 $\mu$m H$_2$O features.}
\label{fig:max_amp}
\end{figure*}

\begin{figure*}
\centering
\begin{minipage}{1\textwidth}
  \centering
  \includegraphics[width=1\linewidth]{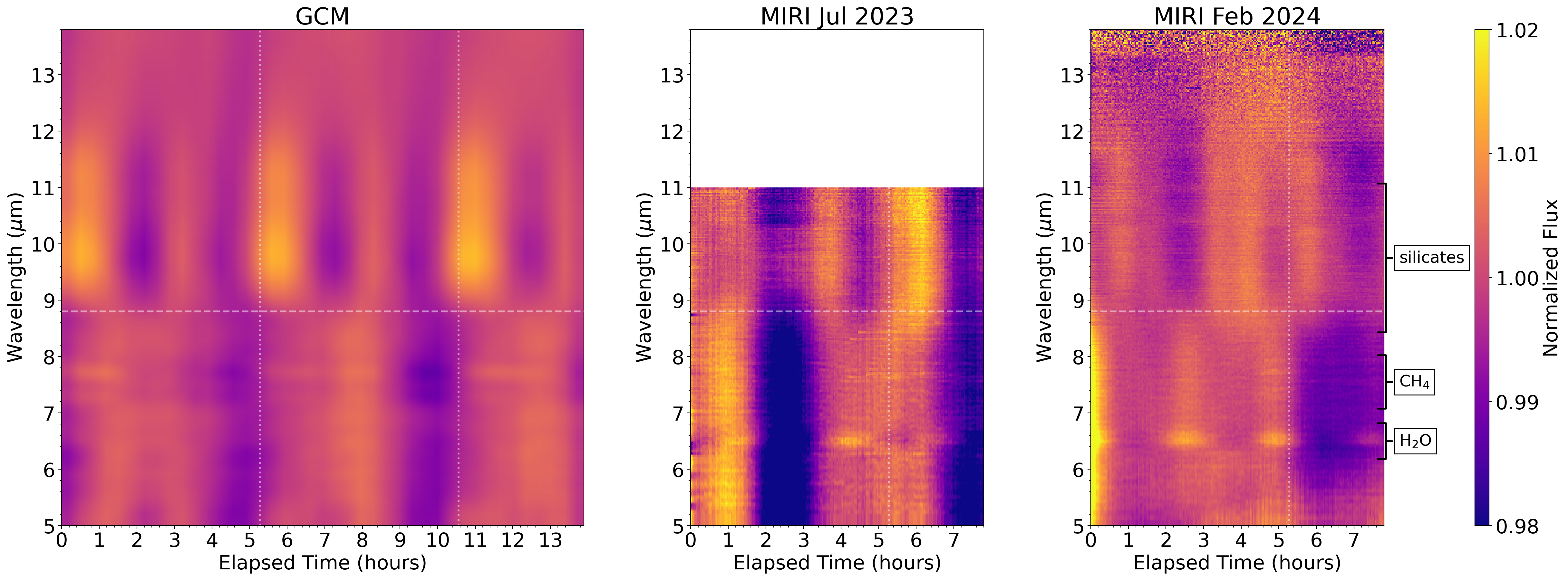}
  \vspace{2pt}
\end{minipage}
\begin{minipage}{1\textwidth}
  \centering
  \includegraphics[width=1\linewidth]{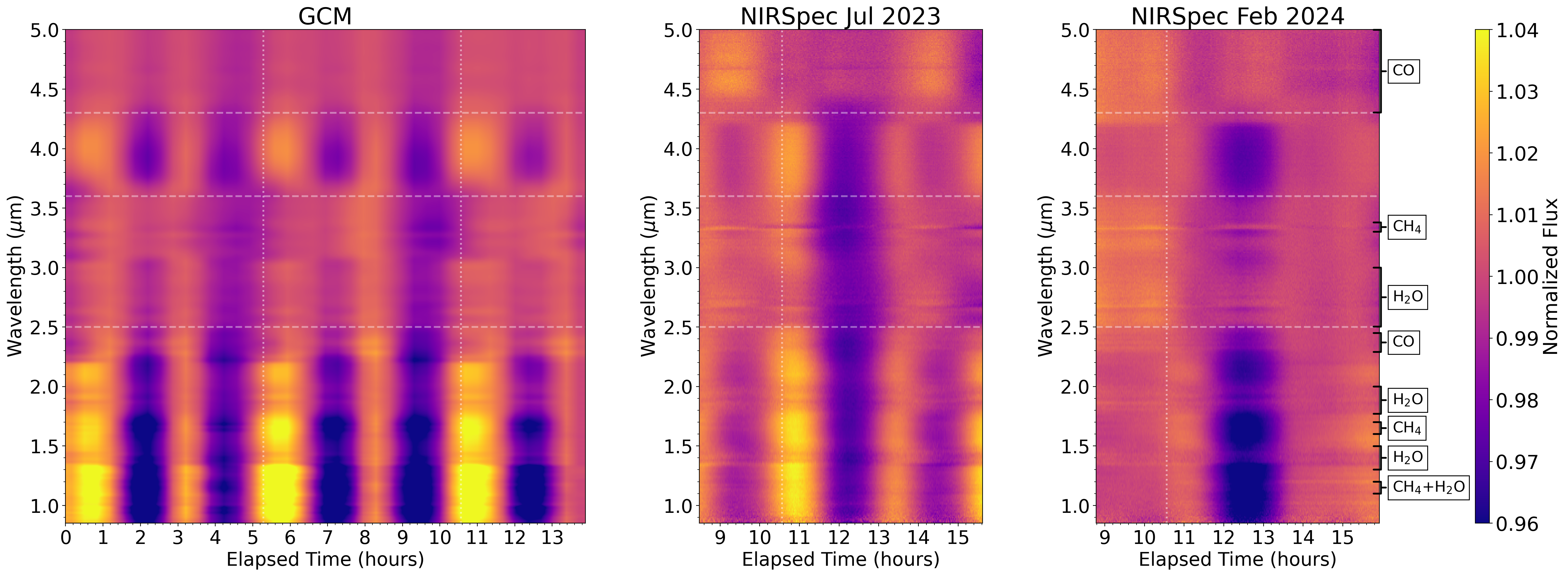}
\end{minipage}
\caption{Variability maps of WISE 1049B predicted by GCM vs. variability maps observed by {\sl JWST} in the July 2023 epoch \citep{Biller2024} and the Feb 2024 epoch (this paper). Vertical dotted lines mark the ends of one rotation period of WISE 1049B adopted in the GCM (5.28 h) and horizontal dashed lines denote the wavelength of main transition points where light curves change their behaviour and are assigned to a different cluster. Note than we do not attempt to align the rotational phase of the GCM and the {\sl JWST} observations, therefore the vertical lines are only indicative for the timescales. The GCM reproduces major transitions
of light-curve behaviours over wavelengths observed in both epochs of WISE 1049B. }
\label{fig:gcm_varmap}
\end{figure*}

Following the indicative comparisons with atmospheric models in the previous section, we now look at how our observed variability compares to a general circulation model (GCM) with a more energetically and dynamically self-consistent modeled inhomogeneous atmosphere in 3-D, further examining the key variability mechanisms via direct comparisons. In this subsection, we briefly introduce the GCM and the mechanism of cloud radiative feedback in driving the variability, then focus on comparing two aspects: 1) variability amplitude as a function of wavelength and 2) the qualitative light curve behaviour transitions over wavelength. We included both epochs of data from July 2023 \citep{Biller2024} and Feb 2024 (this paper) to examine the long-term variability trends.

In the GCM used in this paper, we solve the global primitive equations of dynamical meteorology that govern the dynamics of the outer atmospheric layers of giant planets and brown dwarfs based on the MITgcm \citep{adcroft2004,showman2009}. The GCM uses a non-grey radiative transfer module \citep{marley1999,showman2009} to calculate realistic heating/cooling rates and thermal structures and assumes equilibrium chemistry with a solar atmospheric composition. We include tracer equations to model the formation, evaporation, and sedimentation of MgSiO$_3$, Fe, and Na$_2$S cloud cycles. Radiative feedback of clouds in the radiative transfer and dynamics is implemented and is the key driver of the circulation and large-scale inhomogeneities. This GCM version updates from previous work of \cite{Tan2021,Tan2021b} that used a simpler grey radiative transfer. A similar model was presented in our earlier {\sl JWST} observation work of WISE 1049AB \citep{Biller2024} and will be fully described in a coming work (Tan et al., submitted). After the mean state of the GCM reached a statistical equilibrium, we post-process the time-dependent 3-D temperature and cloud structures using \textsc{picaso} \citep{Batalha2019, Mukherjee2023} to generate spectra and spectroscopic light curves over 2 rotations of the WISE 1049B model assuming an equator-on viewing angle (as suggested by high-resolution spectroscopic observations, e.g., \citealp{Chen2024}) and derived the maximum deviation of normalized light curves in the same way as for the observed light curves (see below).  Parameters of WISE 1049B are adopted as the gravity of 588 ms$^{-2}$, radius of 1.05 $R_\mathrm{Jup}$ and a rotation period of 5.28 h.

\begin{figure*}
\includegraphics[width=1\textwidth]{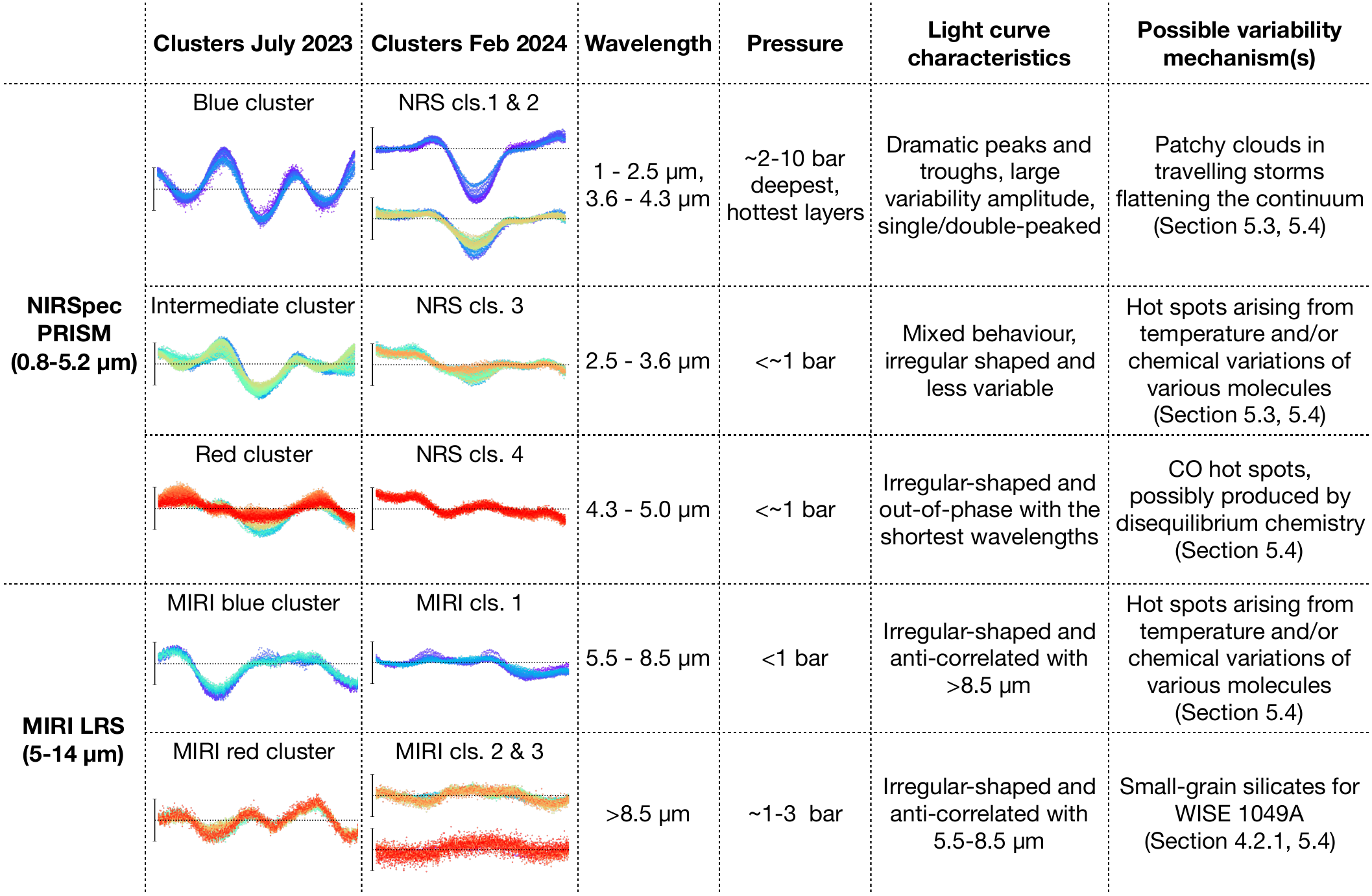}
\caption{Summary of WISE 1049AB's atmosphere layers identified from the NIRSpec and MIRI light curves clusters in the July 2023 and Feb 2024 epoch, and the proposed variability-driving mechanisms for each layer. The light curve clusters are normalized individually with respect to each of their median value and plotted on the same y-axis scale. The black vertical bars on the left indicates a flux variation of $\pm$2\% for reference. The section numbers where each mechanism is discussed are included for reader's reference.}
\label{fig:table}
\end{figure*}

The light curve variability predicted by this GCM model is primarily driven by cloud radiative feedback. When clouds form in a certain area, the greenhouse warming by the cloud opacity exerts heating in that area, whereas in relatively cloud-free areas flux can escape from deeper layers and cool the local atmospheric column. The horizontal heating and cooling contrast drives a large-scale circulation that maintains and triggers cloud cycles. Due to the planetary rotation, vortices form at high latitudes with chaotic evolution \citep{Tan2021b} and large-scale, east-west traveling tropical storms form at low latitudes \citep{Tan2021}. In a nearly equator-on viewing angle, the tropical storms and the associated cloud and temperature inhomogeneity are the main cause of rotational light-curve modulation. The resulting circulation is a chaotic evolving system, meaning that the morphologies of tropical storms and vortices do not stay fixed but change over time, much like the mid-latitude meteorological system in the Earth's atmosphere. Therefore, propagation, chaotic dissipation and resurgence of tropical storms shape the irregular evolution of light curves (see a data-GCM comparison in the long-term TESS light-curve analysis of \citealp{Fuda2024}). 
These tropical storms sometimes are dominated by a single wave packet across the equator, corresponding to one sinusoidal light curve over one rotation period; but sometimes contain two wave packets across the equator, corresponding to two sinusoid-like features over one rotation period. This mechanism could explain the transition of the 1--2.5 $\mu$m light curves from being double-peaked in the July 2023 epoch to single-peaked in this 2024 epoch (see the summary in Fig. \ref{fig:table} for a direct comparison).
Note that, despite clouds being the root causing, significant temperature and chemical anomalies (the latter comes from either chemical equilibrium or disequilibrium chemical mixing associated with cloud dynamics, \citealp{Lee2024}) co-exist and contribute to the variability.

We now examine the overall variability amplitude predicted by the GCM compared to our data. Because the light curves deviate significantly from a simple sinusoidal shape, standard methods of estimating variability amplitude by fitting sinusoidal curves are not suitable for our light curves. Instead, we quantify the amplitude of variability using the maximum deviation. 
We define the maximum deviation for the light curve at each wavelength as the difference between the highest and lowest flux observed throughout each 7–8 h MIRI or NIRSpec observation, then normalized by the time-averaged flux to highlight the change in percentage, i.e., at each wavelength point, maximum deviation ($\lambda$) = $(F_{\rm{max, \lambda}} - 
F_{\rm{min,\lambda}} )/ F_{\rm{median, \lambda}}$. 
Note that with this definition, the maximum deviation is twice the deviation from the mean spectrum, which is similar to the typical variability amplitude reported in the literature.
Fig. \ref{fig:max_amp} shows the maximum deviation as a function of wavelength for this epoch and the July 2023 epoch, compared to predictions from the GCM.

The GCM predicts that variability is highest in the near-IR and decreases to remain relatively flat at $\sim$2\% level in the mid-IR, until another slight peak is seen around 9--10 $\mu$m. Compared to the GCM prediction, the observed maximum deviation of WISE 1049B follows a similar trend, with variability generally higher in the near-IR than in the mid-IR.
The predicted variability is the highest in the continuum in between the H$_2$O absorption bands from 1--2.5 $\mu$m, as well as in the continuum between 3.5--4 $\mu$m. The peak in maximum deviation of WISE 1049B at these wavelengths is consistent with the predictions. Since there are no molecular absorption bands in these ranges and the spectra is probing the continuum, this could indicate the effect of patchy clouds on modulating the flux from the continuum, as these cloud structures rotate in and out of view over time.
Additionally, peaks of variability around the 3.3 $\mu$m CH$_4$ feature are correctly predicted by the GCM. We also observed peaks much higher than those predicted by the GCM, which are correlated with spectral absorption features at the 2.7 $\mu$m H$_2$O, 4.5 $\mu$m CO, and 6.5 $\mu$m H$_2$O. However, the GCM in general overpredicts the amplitudes at short wavelengths and underestimates the amplitudes at long wavelengths, or equivalently, predicts a steeper amplitude trend as a function of wavelength, than that observed, a similar issue shown in the data-GCM comparisons of \cite{Biller2024}.

For WISE 1049A, its near-IR variability is much flatter. Since the GCM model is tuned to the parameters for WISE 1049B, the variability signatures of WISE 1049A could be different from the model simply due to it's different atmospheric structures (clouds, gases, temperatures, etc). Interestingly, the peak in maximum deviation at 8.5--11 $\mu$m is similar to that predicted by the GCM at 9-11.5 $\mu$m, but only slightly offset in wavelength. This is likely a result of a small fraction of small MgSiO$_3$ particles (with a mean radius=0.7 $\mu$m) implemented in the GCM. The offset of the start of the bump at 8.5 vs. 9.0 $\mu$m could point to a slightly different silicate composition from the GCM. The fact that this bump in variability at the silicate absorption feature is seen in WISE 1049A but not in WISE 1049B is consistent with this feature being weakly detected in A's spectrum but undetected in B's. 
Beyond 12 $\mu$m, the rise in observed variability is likely due to poor SNR.

Now we turn our attention to comparing the qualitative light-curve-behaviour transitions as a function of wavelength. Fig.~\ref{fig:gcm_varmap} shows the variability maps predicted by GCM compared to variability maps of WISE 1049B observed by {\sl JWST} in both epochs, as presented earlier in Section~\ref{sec:varmap}. The variability maps produced by the GCM showed interesting light curve behaviour transitions at several wavelengths that roughly match the observed WISE 1049B light curves. The GCM correctly produced transitions at around 2.5, 3.6, and 4.3 $\mu$m in the NIRSpec wavelengths, and around 8.5 $\mu$m in the MIRI wavelengths. 
The regions of $<$2.5 $\mu$m and 3.6--4.3 $\mu$m are in phase, whereas 2.5--3.6 $\mu$m are out of phase, similar to the observed ones. The GCM has more distinctive light curves than the observed ones in CH$_4$-dominated bands such as 3.1--3.6 $\mu$m and 7.2-7.8 $\mu$m. 
The main distinction between GCM and observation happens at 4.3--5.0 $\mu$m, where GCM is in phase with the continuum whereas observed light curves are out of phase. It is possible that at $>$4.3 $\mu$m, the GCM is missing enough CO because of the equilibrium chemistry assumption. In reality, the carbon species including CO and CH$_4$ may be quenched by dynamics (and therefore less sensitive to horizontal temperature variations) and responsible for the out-of-phase behaviour at $>$4.3 $\mu$m as well a less distinct light-curve behaviour near the CH$_4$ bands for the observed light curves. Further GCM efforts incorporating disequilibrium chemistry will be needed to understand the variability at these wavelengths \citep{Lee2023, Lee2024}.

In summary, the GCM qualitatively captures the general trend of light-curve amplitudes as a function of wavelength but this trend is quantitatively steeper than that observed (Fig. \ref{fig:max_amp}). In comparisons of the variability maps (Fig.~\ref{fig:gcm_varmap}), the GCM reproduces major transitions of light-curve behaviours over wavelengths observed in both epochs of WISE 1049B. 
These results, combined with previous discussion in Sections \ref{sec:cf} and \ref{sec:devspectra}, depicts a 3-D picture of the atmosphere of WISE 1049B (summarized in Fig. \ref{fig:table}): 
In the deepest and hottest layer probed by wavelengths of 1--2.5 $\mu$m and 3.6--4.3 $\mu$m, light curves emerge from the continuum regions and are preferentially shaped by the the vertically coherent cloud structures  rotating in and out of view. This produces dramatic light curves with large peaks or troughs, as observed in the NIRSpec blue cluster in the 2023 epoch and NIRSpec Cluster 1 \& 2 in the 2024 epoch. 
Higher up in the atmosphere, light curves emitted from strong gas opacity bands are more irregular shaped and out-of-phase with deeper levels. This layer is probed by NIRSpec Cluster 3 \& 4 covering H$_2$O and CH$_4$ absorption bands at 2.5--3.6 $\mu$m and CO band at 4.3--5.0 $\mu$m, and by MIRI Cluster 1 (MIRI blue cluster in 2023) at 5--8.5 $\mu$m. The light curve behaviours at this level are likely driven by hot spots arising from horizontal temperature and/or chemical variations of various molecules (e.g. H$_2$O and CH$_4$). 
Finally, at wavelengths longer than 8.5 $\mu$m, there is an intermediate pressure level around 1--3 bar, where small-grain silicates potentially contributed to WISE 1049A's variability at 8.5--11 $\mu$m. This level includes MIRI Cluster 2 \& 3 in this epoch and the MIRI red cluster in 2023. The characteristic light curves of each cluster / atmosphere level and their identified possible variability mechanisms are summarized in Fig. \ref{fig:table}. These findings highlight the interplay of cloud and chemistry in shaping the complex atmosphere of WISE 1049AB. While different atmospheric layers are likely governed by different variability mechanisms, each mechanism appears to remain the same within its respective layer over the long term.

\subsection{Comparison with other objects}

The only other brown dwarf/planetary-mass object with published 1--12 $\mu$m {\sl JWST} spectroscopic monitoring is the T2.5 brown dwarf SIMP 0136+0933 \citep{McCarthy2024}, which has similar $T_\mathrm{eff}$ at 1100--1200 K as WISE 1049AB, but lower mass ($\sim$13 $M_\mathrm{Jup}$), considerably younger age (200 Myr) and lower surface density ($\log g$=4.31$\pm$0.03) \citep{Gagne2017}. Some other key differences from the WISE 1049AB system include SIMP 0136's rapid rotation, with a period of 2.4 hours \citep{Artigau2009}, and the presence of radio aurorae \citep{Kao2016}. Compared to WISE 1049AB, the light curves from SIMP 1036 displayed less period-by-period evolution in the NIRSpec-MIRI overlapping wavelength regions around 5 $\mu$m. By linking light curve clusters to contribution functions and comparing deviation spectra with models, \cite{McCarthy2024} found that the SIMP 0136 light curves at the shortest NIRSpec wavelengths ($\sim$0.8–2.2 $\mu$m) are influenced by clouds, while hot spots—possibly caused by aurorae, rising or falling air pockets, or temperature perturbations—likely affect wavelengths in the $\sim$2.2–3.7 $\mu$m range.
SIMP 0136 shows a similar transition of light curve behaviours at 8.5 $\mu$m, coinciding with the end of hot spot-driven variations and the onset of the silicate absorption feature. 
This shared change of behaviour around $\sim$8.5 $\mu$m indicates a common underlying mechanism at these wavelengths for L-T type brown dwarfs like SIMP 0136 and WISE 1049AB. 
Although the 8--11 $\mu$m small-grain silicate absorption feature is not directly observed in the MIRI spectrum of SIMP 0136 and WISE 1049B, atmospheric retrievals (e.g. \citealt{Vos2023}) have favored silicate cloud models for SIMP 0136 over all other models. This suggests that silicate clouds may be just barely visible in the mid-IR photosphere, while the clouds driving the dramatic light curves at shorter wavelengths ($<$2.5 $\mu$m) may reside in the near-IR photosphere.
A dedicated comparison study would improve our understanding of the differences between atmospheres of various spectral types, and spectroscopic variability observations of more targets along the L-T-Y spectral sequence with {\sl JWST} will allow us to trace and confirm the variability mechanisms across a wider population of brown dwarfs and planetary-mass objects.

\section{Conclusions}
\label{sec:conclusion}

In this paper, we report results from a second epoch of {\sl JWST} MIRI LRS + NIRSpec BOTS PRISM variability monitoring of the benchmark brown dwarf binary WISE 1049AB, observed in February 2024. This is the first variability monitoring study of any brown dwarf and planetary-mass object study to measure variability beyond 11 $\mu$m, reaching up to 14 $\mu$m. Our dataset, combined with the first epoch of observation in July 2023, also sets the longest baseline for {\sl JWST} weather monitoring to date, allowing us to robustly test variability-driving mechanisms over the long term for the first time. We summarize our key conclusions below:

\begin{enumerate}
    \item We obtained resolved NIRSpec and MIRI traces for both components of the binary and extracted spectral time series from 1 to up to 14 $\mu$m using multiple spectral extraction methods, including PSF-fitting. The light curves produced from all extraction methods are consistent. We found several H$_2$O, CH$_4$, and CO absorption features in the NIRSpec and MIRI spectra for both components of the binary. The $\sim$9 $\mu$m silicate absorption feature is weakly detected in the MIRI spectra of WISE 1049A, but not detected in WISE 1049B.
    
    \item We constructed narrowband light curves from 1--14 $\mu$m, which show that both components are variable at all wavelengths observed. We found that WISE 1049B is more variable than A in the near-IR whereas WISE 1049A is more variable than B in the mid-IR during this epoch. The light curves show complex, wavelength-dependent behaviours. WISE 1049B display a dramatic trough at 3.8 h in the light curves between 1.0--2.5 $\mu$m and 3.6--4.3 $\mu$m, and the light curves at 4.3--5.0 $\mu$m are out-of-phase with the shorter wavelengths. Both WISE 1049 A and B show out-of-phase light curves below vs. beyond 8.5 $\mu$m.
    
    \item Combining $\sim$15 h of MIRI + NIRSpec light curves in the overlapping wavelengths from 5--5.2 $\mu$m, we found period-to-period light curve variations for both components. Using a Lomb–Scargle periodogram analysis, we found a $\sim$6.7 h period for WISE 1049A and a $\sim$5 h period for WISE 1049B, which are consistent with previous findings.
    
    \item We conducted a cross-epoch absolute flux comparison with the July 2023 epoch and estimated the bolometric luminosity of WISE 1049AB in the 2024 epoch. We found flux variations in the NIRSpec and MIRI broadbands individually, but we did not found significant changes in the bolometric luminosity estimated from the JWST bandpass for both components. 
    
    \item We performed k-means clustering analysis for the 2024 epoch spectroscopic light curves. For WISE 1049B, we found 4 different characteristic light curve clusters in the NIRSpec wavelengths and 3 different clusters in the MIRI wavelengths. For WISE 1049A, we found 3 light curve clusters from NIRSpec and 3 clusters from MIRI. Comparing them to a contribution function from a Diamondback model at 1300K with $f_\mathrm{sed}$=8 clouds, we found that the WISE 1049B clusters roughly correspond to 3 distinctive pressure layers in its atmosphere: 1) A deep and hot layer probed by NIRSpec clusters 1 \& 2 at 1--2.5 $\mu$m and 3.6--4.3 $\mu$m; 2) a high-altitude layer probed by NIRSpec cluster 3 \& 4 and MIRI cluster 1 from 2.5--3.6 $\mu$m and 4.3--8.5 $\mu$m; and 3) an intermediate layer probed by MIRI cluster 2 \& 3 at the longest wavelengths $>$8.5 $\mu$m. We found that the wavelengths at which cluster transitions occur are broadly consistent with the previous epoch, suggesting consistency in the variability mechanisms responsible for each cluster over timescale of several months.
    
    \item By looking at the deviation from the mean spectra at different rotational phases of WISE 1049B, and comparing to a general circulation model (GCM), we identified the possible mechanisms that drive the light curves at these various pressure levels: Patchy clouds rotating in and out of view likely shaped the dramatic light curves with deep troughs in the deepest layers $<$2.5~$\mu$m, whereas CO and CH$_4$ hot spots originating from vertical mixing and temperature / chemical variations likely dominate the high-altitude levels between 2.5--3.6 $\mu$m and 4.3--8.5 $\mu$m. Small-grain silicates potentially contributed to the variability of WISE 1049A at 8.5-11 $\mu$m. 
    Building on robust clustering analysis and data-GCM comparison on the two epochs, we conclude that while distinct atmospheric layers are likely governed by different variability mechanisms, each mechanism appears to remain consistent within its respective layer over the long term.

\end{enumerate}

Consecutive multi-period observations in the future will further confirm the nature of period-to-period and long-term evolutions in WISE 1049AB's light curve behaviour and verify the stability of their variability mechanisms. Robust variability monitoring of more brown dwarfs, planetary-mass objects, and directly imaged companions in the next few {\sl JWST} cycles will transform our understanding of these extrasolar atmospheres.

\section*{Acknowledgements}

XC acknowledges funding from the China Scholarship Council (CSC) under Grant CSC No. 202208170018. 
BAB and BJS acknowledge funding by the UK Science and Technology Facilities Council (STFC) grant no. ST/V000594/1. 
XT is supported by the National Natural Science Foundation of China (grant No. 42475131),.
JMV acknowledges support from a Royal Society - Research Ireland University Research Fellowship (URF/1/221932). 
AMM acknowledges support from the National Science Foundation Graduate Research Fellowship under grant no. DGE-1840990. 
NOG acknowledges the support from the Arthur Davidsen Graduate Student Research Fellowship \textit{provided by} the Space Telescope Science Institute.
XC and BAB thanks Sarah Kendrew and Greg Sloan for providing an MIRI PSF built from commissioning data and for advice on how to implement PSF-fitting for MIRI LRS data.

\section*{Data Availability}

All raw and pipeline-processed data presented in this article are available via the MAST archive. After publication, we will provide the custom-reduced data products described in this article on Zenodo at doi: 10.5281/zenodo.12531991.



\bibliographystyle{mnras}
\bibliography{references} 




\appendix

\section{Row-by-row light curves from NIRSpec}

\begin{figure*}
\includegraphics[width=1\textwidth]{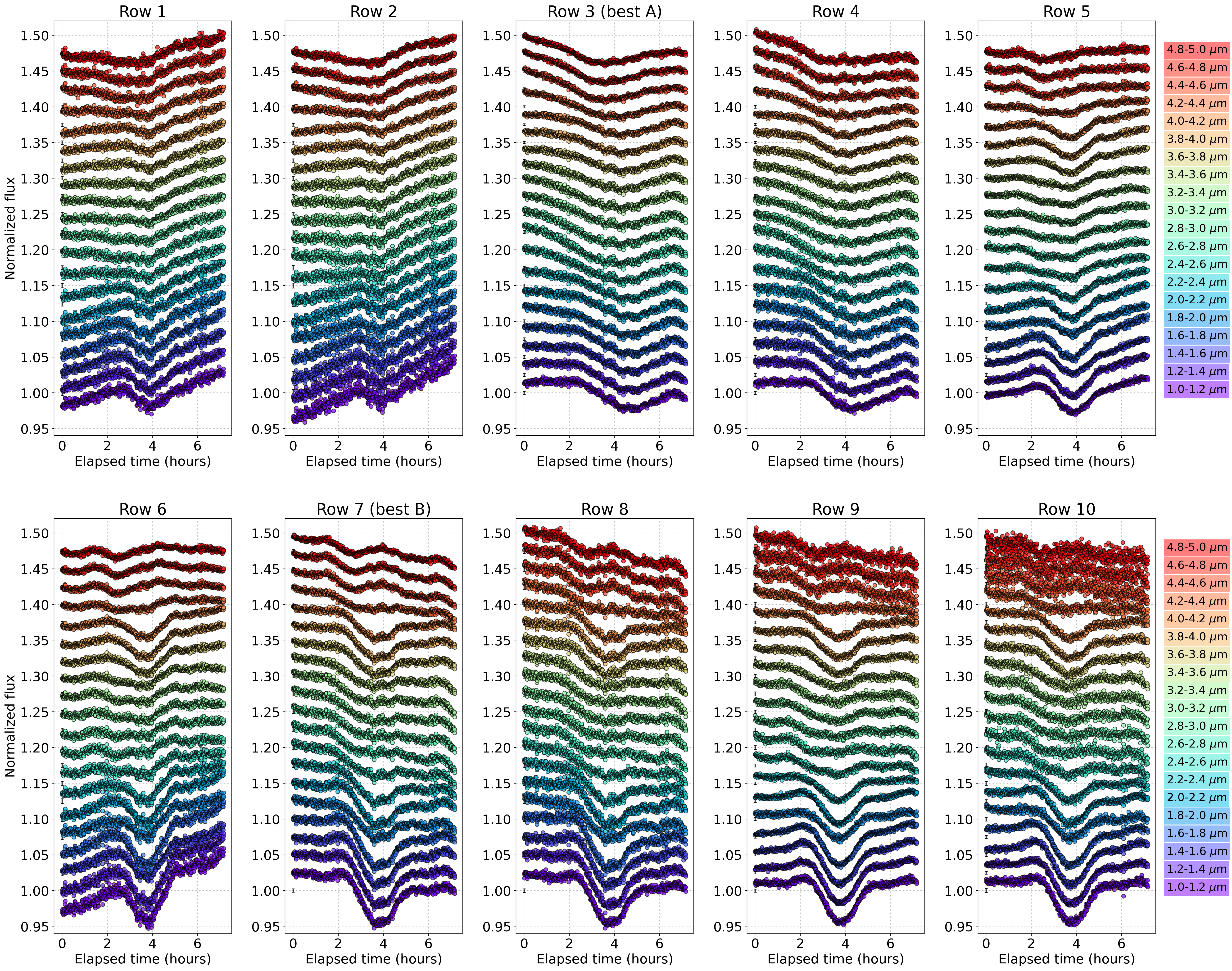}
\caption{NIRSpec WISE 1049AB light curves extracted from each row on the detector. Row 3 and row 7 represent the least contaminated rows on the detector where traces for WISE 1049A and B can be found respectively. The more chaotic or noisy light curves on the other rows is a result of blending between the original PSF of WISE 1049A or B with the extra PSF(s) from the other component produced by the {\sl JWST} mirror tilt event.}
\label{fig:rowbyrowlc}
\end{figure*}


\bsp	
\label{lastpage}
\end{document}